\documentclass[graybox, envcountchap]{svmult}

\usepackage{mathptmx}        
\usepackage{amsmath}
\usepackage{amssymb}
\usepackage{color}
\usepackage{helvet}          
\usepackage{courier}         
\usepackage{dirtree}

\usepackage{makeidx}        
\usepackage{graphicx}        
\usepackage{subfig}

\usepackage{multicol}        
\usepackage[bottom]{footmisc}

\usepackage{hyperref}        
\hypersetup{colorlinks=true,urlcolor=blue}

\usepackage[misc]{ifsym}

\usepackage{aas_macros} 
\usepackage{xspace}
\newcommand {\ixpe}{{IXPE}\xspace}

\newcommand{\degr}{$^\circ$\xspace}

\makeindex             

\begin{document}


\title{The hitchhiker's guide to the IXPE data analysis}

\author{Alessandro Di Marco}

\institute{Alessandro Di Marco \at INAF-IAPS, via Fosso del cavaliere 100, I-00133, Rome, Italy, \email{alessandro.dimarco@inaf.it}}
%
%
\maketitle

\abstract{This chapter provides an almost comprehensive guide to the Imaging X-ray Polarimetry Explorer (\ixpe) data analysis. The chapter briefly introduces the \ixpe spacecraft and the instrument onboard; subsequently, the strategies adopted to extract polarimetric information and to optimize the response are given. Moreover, the data formats and processing steps to avoid potential systematic errors and to achieve the best results from the \ixpe data are reported. Both the model-independent and the model-dependent analyses are summarized, as well as the instrument response functions and their different available flavors. The aim of this chapter is to compile suggestions and answers to questions commonly raised by users, to help researchers maximize the scientific return from \ixpe observations.}


\section{Introduction: The Imaging X-ray Polarimetry Explorer}
\label{sec:intro}

The Imaging X-ray Polarimetry Explorer (\ixpe), launched on 9 December 2021, is a NASA Small Explorer mission in collaboration with the Italian Space Agency (ASI), fully dedicated to X-ray polarimetry \cite{Soffitta2021,Weisskopf2022}. \ixpe has on board three identical co-aligned telescopes; each telescope consists of an X-ray mirror-module-assembly (MMA) having a $\sim$4\,m focal length and at the focus a polarization-sensitive detector, the Gas Pixel Detector (GPD) \cite{Baldini21}, providing imaging, timing, and spectral information within the 2--8\,keV energy range in addition to polarization. The mirrors and detectors are separated and aligned by a boom. The three \ixpe focal plane detector units (DUs) are clocked 120\degr apart to mitigate detector-correlated residual effects in the determination of polarization. 

Each DU hosts a GPD, having a $15\times15$\,mm$^2$ sensitive area. These detectors can detect X-ray polarization via the photoelectric effect, the primary process for X-ray photons. The photoelectric effect is due to the absorption of a photon from an atomic electron, mainly in the K shell, followed by the emission of a free electron (hereafter named photoelectron); when X-rays have the electric field vector along a preferential direction, $\phi_0$, they are polarized, and the photoelectron emission preferentially aligns to such a direction following the cross-section equation \cite{Heitler1954}:
\begin{equation}
    \frac{d\sigma_{\rm{K}}}{d\Omega}\propto \frac{\sin^2 \theta \cos^2 (\phi-\phi_0)}{(1+\beta \cos^4 \theta)},
\end{equation}
where $d\sigma_{\rm{K}}/d\Omega$ is the differential cross section with respect to the solid angle $\Omega$, $\phi$ and $\theta$ represent the azimuthal and polar angles of the photoelectron ejection direction, respectively, and $\beta$ is the electron velocity in units of the speed of light. Thus, the ejection direction of the photoelectrons exhibits an angular distribution modulated by ${\propto}\cos^2 (\phi - \phi_0)$, whose amplitude and phase provide information on the degree and angle of polarization of the detected X-rays \cite{DiMarco22b}.

The GPD's working principle is similar to traditional proportional counters' by using a gas mixture to capture incoming X-rays, alongside an applied voltage that drifts the electron charge cloud towards an anode readout; the difference with respect to proportional counters is the presence of a gas electron multiplier to facilitate charge gain, maintaining the photoelectron track shape. The anode readout of GPDs is a finely pixelated application-specific integrated circuit (ASIC), which enables effective imaging of photoelectron ionization tracks. For each detected photon, the GPDs provide the photoelectron ejection direction (depending on polarization), its ionization charge, which relates to the photon's energy, the absorption point, and the time of arrival, yielding spectrometry, imaging, and timing measurements (for more details, see Ref.~\cite{Baldini21}).

After a brief commissioning and calibration period, on 11 January 2022, \ixpe began its scientific operations. During a two-year baseline period of {\it prime mission} several serendipitous results were obtained \cite{Prokhorov24,Xie22,Krawczynski22, DiMarco25, LaMonaca24, Baglio25, Doroshenko22, DiMarco25b, Papitto25,Kim24, Ingram23}. Currently, \ixpe has opened its General Observer (GO) program, making its unique capabilities available to the entire scientific community. The following sections report suggestions and guidelines to extract and analyze \ixpe data properly; although this is not an official guide, this is mainly based on the information provided by Refs.~\cite{DiMarco22, DiMarco23a}, the \ixpe\ {\it Quick Start Data Analysis Guide}\footnote{\href{https://ixpe.msfc.nasa.gov/for_scientists/documentation/ixpeQS_v07.pdf}{https://ixpe.msfc.nasa.gov/for\_scientists/documentation/ixpeQS\_v07.pdf}}, AND THE \textsc{ixpeobssim} manual\footnote{\href{https://ixpeobssim.readthedocs.io/en/latest/index.html}{https://ixpeobssim.readthedocs.io/en/latest/index.html}} \cite{Baldini22}.

\section{How to measure polarization from photoelectric polarimeters}\label{sec:pol}

\begin{figure}[!hb]
	\centering
	\includegraphics[width=0.48\textwidth]{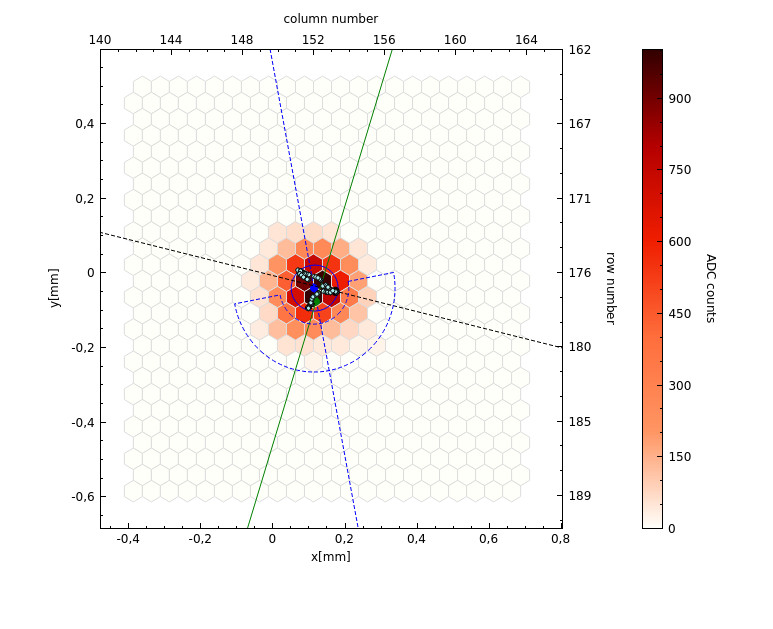}
	\includegraphics[width=0.48\textwidth]{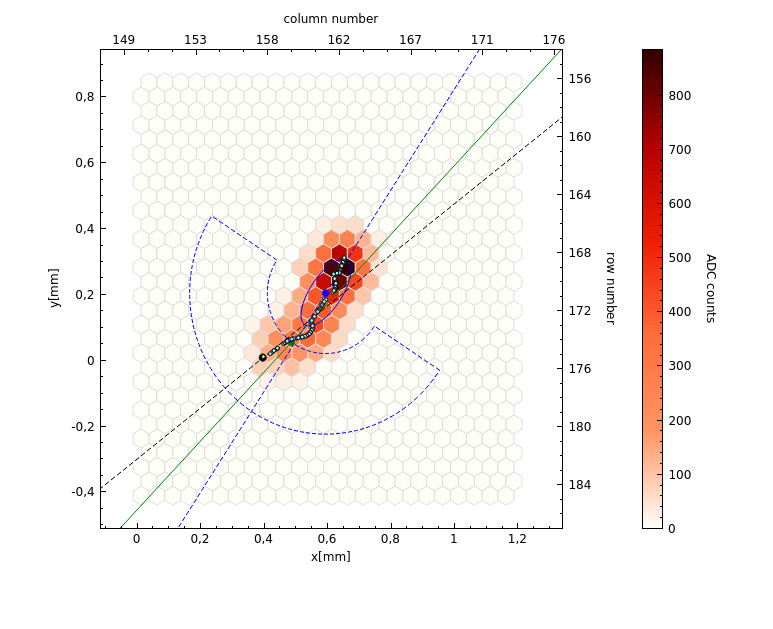}
	\caption{Examples of simulated ionization tracks resulting from the absorption of a 4\,keV X-ray in the GPD. The photoelectron emission directions from the Monte Carlo are shown with a green line, while the blue and black lines show the photoelectron directions reconstructed from the \ixpe algorithm in the first and second steps, respectively. It is evident that the rounded tracks (left) provide a poorer estimate of the direction of photoelectron ejection than the more elongated tracks (right).
	}
	\label{fig:tracks}
\end{figure}

The photoelectron tracks collected by the GPDs on board \ixpe are analyzed by an algorithm (see, e.g., Refs.~\cite {DiMarco22, Bellazzini2003}) to extract relevant information, including the photoelectron track direction, which determines the X-ray polarization. Such an algorithm uses the charge distribution collected in the detected tracks (as shown in Figure~\ref{fig:tracks}) to estimate, in the first step, the barycenter and the second and third moments. Then, in the second step, the final absorption point and the photoelectron's ejection direction are derived. The second step uses the second moment to determine a preliminary (first-step) direction of the track and the third moment to determine the charge asymmetry, thereby identifying the Bragg's peak region. The latter is the region at the end of the photoelectron pattern where most of the energy, and consequently charge, is released. In Figure~\ref{fig:tracks}-right, the darker region due to the Bragg's peak can be observed, and the first-step direction (blue line) relies mainly on these pixels. Because the photon's polarization depends on the initial direction of the photoelectron track before the trajectory is modified by scattering in the GPD's gas, the Bragg's peak region should be excluded from the analysis. To obtain this, the second-step analysis is performed considering only pixels within a horseshoe region, from which a circular region around the barycenter of the charge distribution is removed, as in the ones reported in Figure~\ref{fig:tracks}. This method provides a better estimate of the position where the X-ray is absorbed and of the photoelectron's ejection direction. 

\begin{figure}[!tb]
	\centering
	\includegraphics[width=0.6\textwidth]{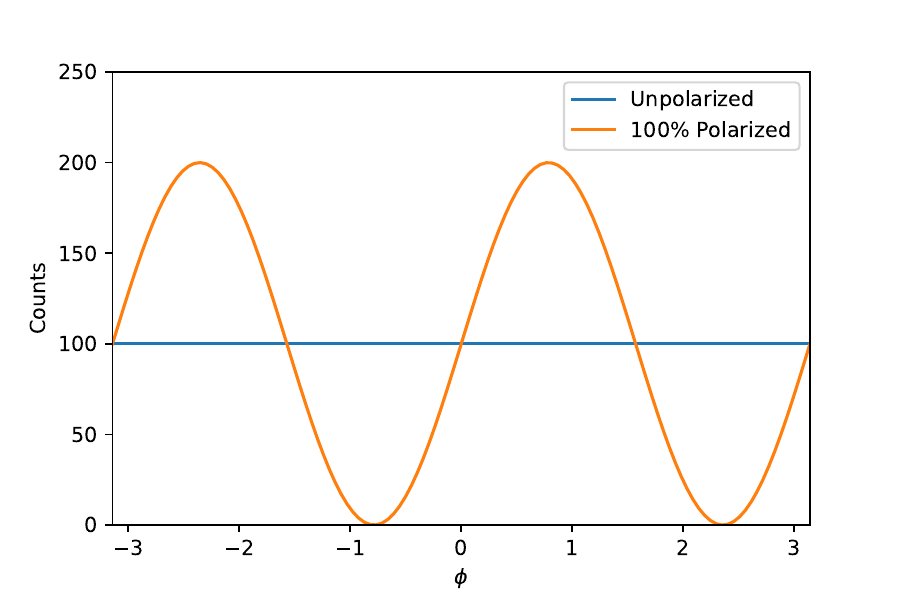}
	\caption{Example of a modulation curve observed with an ideal detector for unpolarized (blue line) and 100\% polarized (orange line) X-rays.}
	\label{fig:modcurve}
\end{figure}

Once the angles at which the photoelectrons are emitted are known, the X-ray polarization can be determined. In fact, as shown in Figure~\ref{fig:modcurve}, the histogram of the photoelectron's azimuthal directions, $\phi$, provides the so-called modulation curve showing different behaviors in case the impinging X-ray source is unpolarized (constant curve) or polarized ($\cos^2\phi$ modulated curve that for 100\% polarized sources ranges from zero up to a maximum of counts).

In the literature, several methods to estimate the polarization degree and angle from the data of photoelectric polarimeters have been proposed: (i) the classical one based on modulation curves fit; (ii) Stokes parameters estimation from modulation curves fit \cite{Strohmayer2013}; (iii) Stokes parameters estimation based on an unbinned event-by-event approach \cite{Kislat15}. These methods, as reported by Ref.~\cite{DiMarco22b}, provide consistent results, but using Stokes parameters offers several advantages. In particular, the ability to treat Stokes parameters as fluxes with full Gaussian statistics allows decoupling of systematic effects in polarization measurements, as described in Ref.~\cite{Rankin22}, or of polarization from different spectral components (see, e.g., Ref.~\cite{Strohmayer2017} and Section~\ref{sec:specpol}). The unbinned approach has the further advantage of allowing estimates to be obtained even from a small number of events and of applying corrections and performing a weighted analysis on an event-by-event basis \cite{DiMarco22}. Hereafter, only this approach is considered because it is the one used for \ixpe data; for each k-th X-ray that produces a photoelectron track in the GPD, the photoelectron azimuthal direction of ejection, $\phi_\textrm{k}$, is determined and translated into Stokes parameters \cite{Kislat15}:
\begin{eqnarray}
i_\textrm{k}&=& 1, \nonumber \\
q_\textrm{k}&=& 2\cos (2\phi_\textrm{k}), \\
u_\textrm{k}&=& 2\sin (2\phi_\textrm{k}). \nonumber 
\end{eqnarray}
After this, a correction is needed because GPDs suffer from a systematic effect known as spurious modulation \cite{Rankin22}. This spurious modulation is an instrumental effect mimicking the polarimetric signal, which can provide a nonzero polarization degree even for unpolarized sources; such an effect has been characterized and is removed thanks to calibration maps depending both on the event's energy, $E_\textrm{k}$, and the absorption point on the GPD sensitive area expressed in pixel coordinates ($x_\textrm{k}, y_\textrm{k}$). The correction is performed by subtracting the effect in the measured Stokes parameters ($i_\textrm{k}$, $q_\textrm{k}$ and $u_\textrm{k}$) on an event-by-event basis \cite{Rankin22}:
\begin{eqnarray}
i_{\textrm{cal},\textrm{k}} & = & i_\textrm{k} ,\nonumber \\
q_{\rm{cal},\textrm{k}} & = & q_\textrm{k} - q_{\rm{sm}}(x_\textrm{k},y_\textrm{k},E_\textrm{k})  ,\\ \nonumber
u_{\rm{cal},\textrm{k}} & = & u_\textrm{k} - u_{\rm{sm}}(x_\textrm{k},y_\textrm{k},E_\textrm{k}) .  
\end{eqnarray}
Here, the subscript ``cal'' denotes the calibrated/corrected values, while the subscript ``sm'' denotes the measured values of the spurious modulation included in the calibration maps. In the end, to obtain an optimal sensitivity for each event, an optimal weight, $w_\textrm{k}$, to the Stokes parameters is applied \cite{DiMarco22, Kislat15}: 
\begin{eqnarray}\label{eq:weights}
I&=&\sum_\textrm{k} w_\textrm{k} i_{\textrm{cal},\textrm{k}}, \nonumber \\
Q&=&\sum_\textrm{k} w_\textrm{k} q_{\textrm{cal},\textrm{k}} , \\
U&=&\sum_\textrm{k} w_\textrm{k} u_{\textrm{cal},\textrm{k}}. \nonumber 
\end{eqnarray}
In the case of GPD's data, the best weight is provided by the track's ellipticity $\alpha = (TL-TW)/(TL+TW)$, with $TL$ the track's length and $TW$ the track's width. In particular, as demonstrated in Ref.~\cite{DiMarco22}, the best polarimetric sensitivity is obtained by using $w_k = \alpha_k^{0.75}$. This weighted analysis requires a statistical approach that differs slightly from the unweighted one, as described in Refs.~\cite{DiMarco22, Kislat15}. Hereafter, the uncertainties and sensitivities are reported for the weighted analysis, but the unweighted case can be obtained simply by setting $w_\textrm{k}=1$. The uncertainties on the normalized Stokes parameters, $q=Q/I$ and $u=U/I$, are given by:
\begin{eqnarray}
\sigma_{q}&=&\sqrt{\frac{W_2}{\mu^2 I^2}\left(2 -\frac{Q^2}{I^2}\right)}, \\
\sigma_{u}&=& \sqrt{\frac{W_2}{\mu^2 I^2}\left(2 -\frac{U^2}{I^2}\right)}, \nonumber
\end{eqnarray}
where $\mu$ is the modulation factor (the measured polarization by a detector in response to a 100\% polarized source) \cite{DiMarco22b}, and $W_2=\sum_{\textrm{k}} w_\textrm{k}^2$. For the unweighted case, $I$ and $W_2$ are both equal to the number of counts, $N$, so the term $W_2/I^2$ is equal to 1/$N$. Because of this, it is possible to define $N_\textrm{eff}=I^2/W_2$. Thus, in the weighted analysis, an effective number of counts, $N_\textrm{eff}$, is used for statistical purposes, instead of the total number of counts, $N$, as in the unweighted approach. $Q$ and $U$ are not fully independent parameters because the polarization degree, $P$, is defined between 0 and 1; thus, a Pearson linear correlation coefficient has to be estimated in addition to $\sigma_q$ and $\sigma_u$ \cite{DiMarco22, Kislat15}:
\begin{equation}
\rho(q,u) = -\frac{\mu^2 P^2\sin(4\phi_0)}{\sqrt{16 - 8 \mu^2 P^2 + \mu^4 P^4\sin^2(2\phi_0)}},
\end{equation}
where $P$ and $\phi_0$ are the polarization degree and the polarization angle:
\begin{eqnarray}
    P&=&\frac{\sqrt{Q^2+U^2}}{I}, \\
    \phi_0 &=& \frac{1}{2}\arctan\left(\frac{U}{Q}\right) . \nonumber
\end{eqnarray}
$\rho$ depends on $P^2$, making it negligible for small degrees of polarization; for polarization angles that are multiples of 45\degr, it is zero, while it is maximized at $-22.5$\degr, where it equals $P^2/3$. This means that even considering the maximum polarization degree $P=1$, this coefficient is $\rho=1/3$, which is considered a moderate correlation. In the end, uncertainties on $P$ and $\phi_0$ can be determined by:
\begin{eqnarray}
\sigma_P & = & \sqrt{\frac{1}{N_{eff}}\left( 1 - \frac{\mu^2 P^2}{2}\right)} , \\
\sigma_{\phi_0} & \approx & \frac{1}{\mu P \sqrt{\frac{1}{N_{eff}-1}}}. \nonumber
\end{eqnarray}
The polarimetric sensitivity is typically determined by using the Minimum Detectable Polarization, MDP$_{99}$, which quantifies the maximum polarization degree that statistical fluctuations can produce for an unpolarized source at a 99\% confidence level; this is defined as \cite{DiMarco22, Kislat15, Weisskopf2010}:
\begin{equation}\label{eq:mdp}
\textrm{MDP}_{99}\approx \frac{4.29}{\mu\sqrt{N_{\textrm{eff}}}}.
\end{equation}

Following the document ``Note on \ixpe Statistics''\footnote{\href{https://heasarc.gsfc.nasa.gov/docs/ixpe/analysis/IXPE\_Stats-Advice.pdf}{https://heasarc.gsfc.nasa.gov/docs/ixpe/analysis/IXPE\_Stats-Advice.pdf}}, the main indicator for the statistical significance of a polarimetric measurement is $P/\sigma_P$. Indeed, because $\chi^2$ is the relevant statistical test, the range of the polarization parameters at a specified confidence level is proportional to $\sigma_P$. A detection of polarization at a given confidence level, $C$, can be claimed when $P/\sigma_P>\sqrt{-2\ln{(1-C)}}$; based on this, if $C>99$\% the detection is ``probable'', it is ``highly probable'' when $C>99.9$\%, while for $C>99.99$\% a ``secure'' detection can be claimed. 
The uncertainties for $P$ and $\phi_0$ reported here do not consider the correlation between these two parameters (1-D errors); to account for this, the best way to report polarization measurements (especially when the result is not highly significant) is by using 2-D confidence regions in the $P$ vs $\phi_0$ plane, called ``protractor plots'', based on $\Delta \chi^2$ levels (suggested confidence levels are 50\%, 90\%, 99\%, 99.9\%).

\begin{figure}[!ht]
\centering
\includegraphics[width=0.9\textwidth]{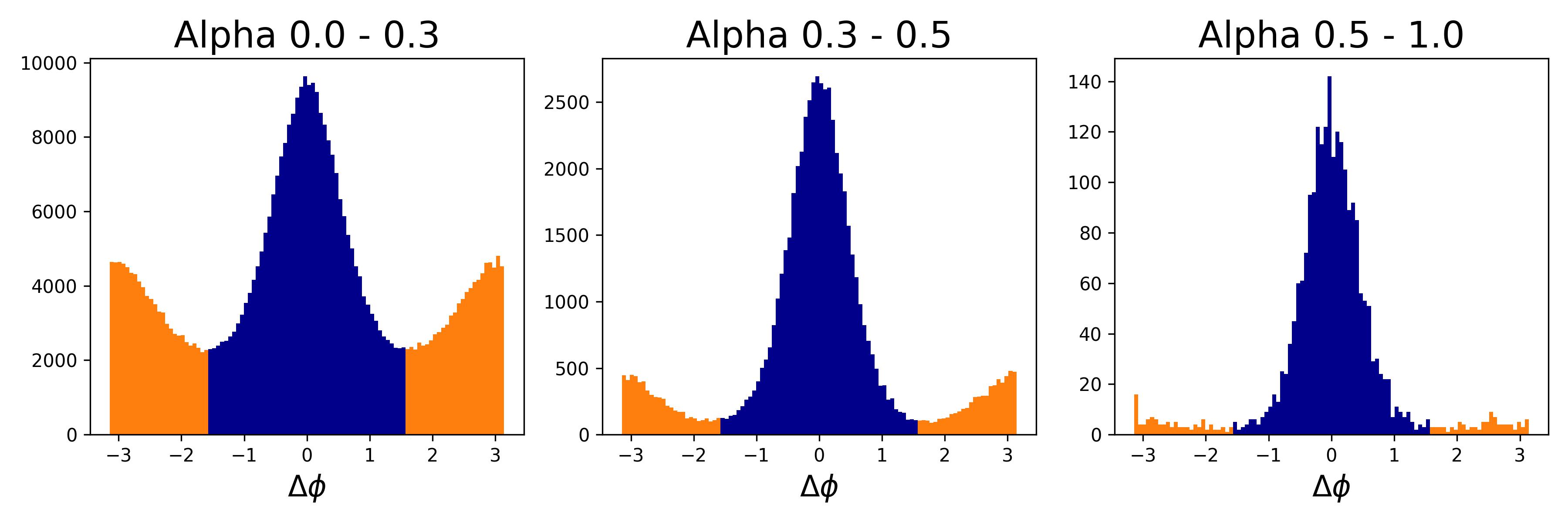}
\includegraphics[width=0.9\textwidth]{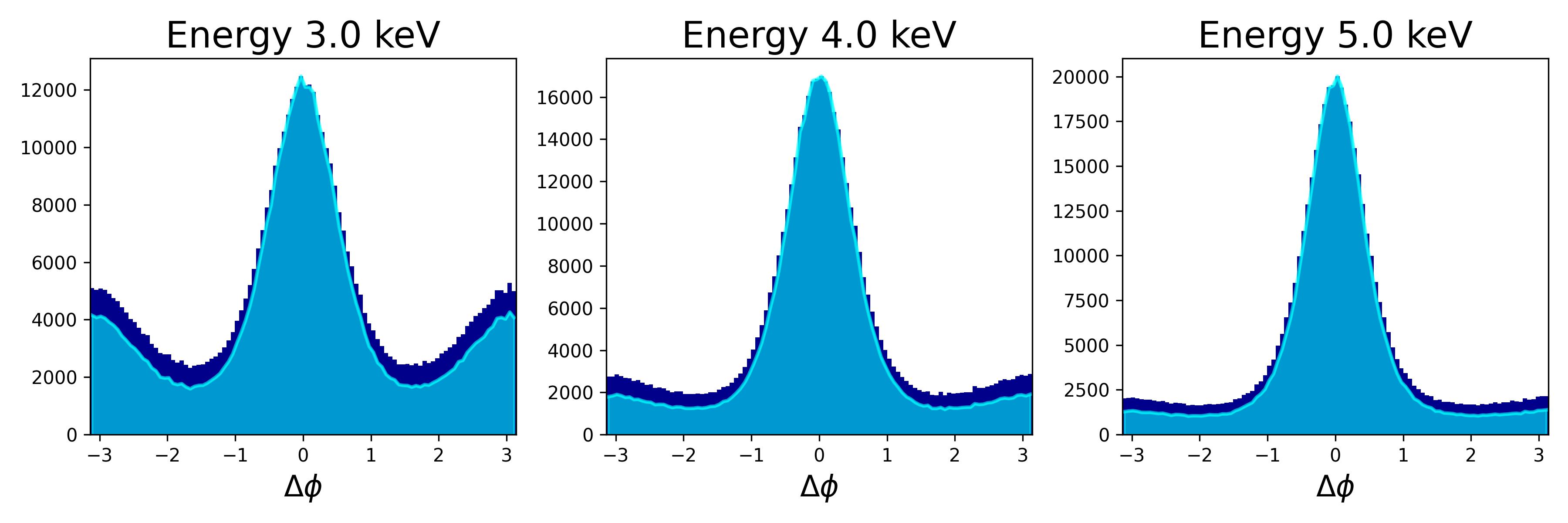}
\caption{(Top) Difference between the reconstructed and true direction angles of emitted photoelectrons, as obtained from simulations, after selecting tracks with different ellipticities; the case where the head and tail of the tracks are inverted is highlighted in orange. (Bottom) Same as in the top panels, but at different energies in the case of unweighted (dark-blue-filled histograms) and weighted (light-blue-filled histograms) tracks.}
\label{fig:weights}   
\end{figure}

The weighted analysis reported here improves the \ixpe sensitivity by ${\sim}13$\% \cite{DiMarco22}. As a further improvement, it allows for the reduction of the ``polarization leakage'' \cite{Bucciantini2023} effect. This effect can lead to radially polarized halos having a polarization degree up to ${\sim}40$\% for isolated point-like sources, but, in general, it is negligible when the region used to select the source is sufficiently large (diameter larger than ${\sim}$30 arcsec, which is the half-energy width) and/or centered on the source. The effect on extended sources depends on the presence of strong gradients in count rates and can be much smaller; the evaluation of the effect in this case can be obtained from simulations \cite{Bucciantini2023}. The ``polarization leakage'' effect is fundamentally related to an incorrect determination of the photon absorption point during track reconstruction and to its correlation with the polarization angle (i.e., when the photoelectrons' emission direction is rotated by 180$^\circ$). As an example, the results from simulations performed with a \ixpe dedicated simulator software assuming a monochromatic source \cite{DiMarco22, Xie21, Kim2024} are reported in Figure~\ref{fig:weights}; in particular, the plots report the difference, $\Delta \phi$, between the true emission direction of the photoelectrons, $\phi_{\rm true}$, as produced by the simulator, and the reconstructed emission direction $\phi$. When $\Delta \phi$ is outside the interval (-90;90), the inversion of the head and tail of the track gives rise to ``polarization leakage'' \cite{Bucciantini2023}. In Figure~\ref{fig:weights}, it is possible to observe that the effect depends on the energy (bottom line) and tracks' ellipticity (top line); this means that the weighted analysis can attenuate the effect of ``polarization leakage'', as demonstrated in Figure~\ref{fig:weights}-bottom, where the inversion probability is reduced thanks to the weighted analysis of ${\sim}15$\% at 3\,keV up to ${\sim}$28\% at 5\,keV.


\section{Data retrieval and formats}
\label{sec:data}

The \ixpe data archive, like those of other NASA high-energy astrophysics missions, is maintained by the High Energy Astrophysics Science Archive Research Center (HEASARC) at the Goddard Space Flight Center (GSFC)\footnote{\href{https://heasarc.gsfc.nasa.gov/}{https://heasarc.gsfc.nasa.gov/}}. Unless otherwise specified, all data files are formatted in the Flexible Image Transport System (FITS), according to HEASARC FITS standards\footnote{\href{https://heasarc.gsfc.nasa.gov/docs/heasarc/fits.html}{https://heasarc.gsfc.nasa.gov/docs/heasarc/fits.html}}. The comprehensive database table that includes all the \ixpe observations that have been archived at the HEASARC is named {\it IXPE Master Catalogue}; in this catalog, an X-ray source, identified by its name, can be present with several table entries corresponding to different observations of the same target. Each observation ID consists of two digits identifying the mission phase or year, a four-digit target ID for that mission phase, and a two-digit segment number. The latter is 01 for the first or unique pointing and 99 when multiple pointings have been merged into the same observation ID, while 02, 03, etc. are used for multiple pointings of the same source in that mission phase. For example, observation ID 04-2508-01 indicates the target 2508 during the 4$^\textrm{th}$ year of observations with a single pointing.

\begin{figure}[!ht]
	\centering
	\includegraphics[width=0.5\textwidth]{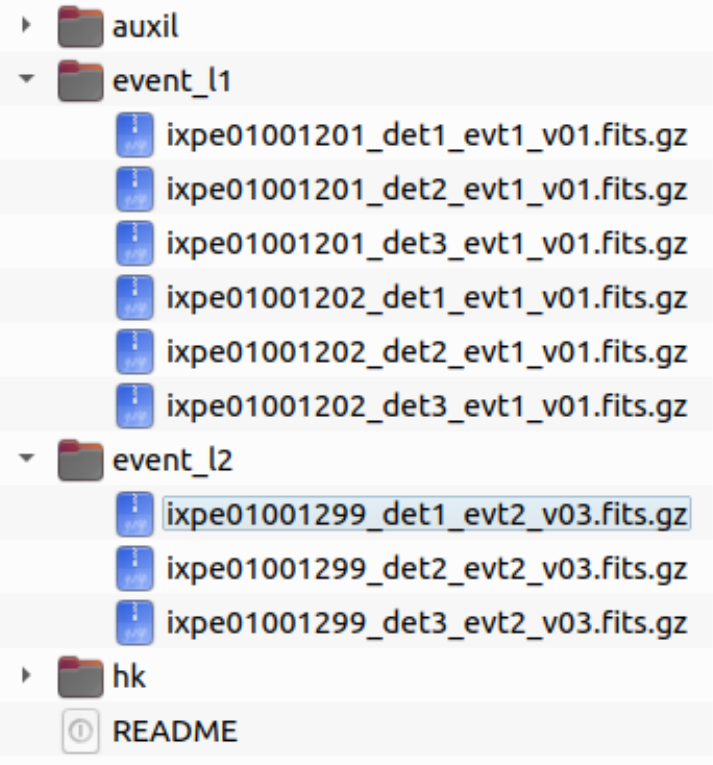}
	\caption{Example of the content from one of the \ixpe observations. It is possible to notice the different folders, the presence of a README file reporting issues, and three Level-2 files (\texttt{evt2}) obtained from six Level-1 files (\texttt{evt1}). The Level-2 data are obtained by merging for each DU the two (one for each pointing) Level-1 files.}
	\label{fig:folder}
\end{figure}

For each observation ID, the \ixpe data are collected in a folder containing housekeeping, Level-1, and Level-2 files (see Figure~\ref{fig:folder}); the latter are required for scientific analysis. 
Some observations may also include a README file reporting possible issues related to that particular observation. In each observation, the files from each \ixpe DU are distributed in different FITS files; thus, three Level-2 files are provided, while Level-1 files may be three or multiples of three if the observation includes more than one pointing.

Level-1 data contains events that are systematically arranged in chronological order on the basis of the events' time recorded by the detector; these timestamps are reported in the \texttt{TIME} column in units of \ixpe Mission Elapsed Time (MET), which is the elapsed time in seconds since January 1$^{st}$, 2017. Raw tracks, preserving information collected by the focal plane detectors, are stored in Level-1 as two-dimensional image data arrays with header terms relevant to the reconstruction algorithm summarized in the previous section. Such an algorithm is implemented in the tool \texttt{ixpeevtrecon}, distributed by HEASoft and included in the \ixpe instrument pipeline used to produce Level-2 files. The users do not need to run it, except in the case they want to reprocess the data by themselves; at this aim, recently the software \texttt{ixpepipe} has been released\footnote{\href{https://heasarc.gsfc.nasa.gov/docs/ixpe/analysis/contributed/ixpepipe.html}{https://heasarc.gsfc.nasa.gov/docs/ixpe/analysis/contributed/ixpepipe.html}}. \texttt{ixpeevtrecon} also provides other parameters that are not included in Level-2, but they are stored in Level-1, such as the position where X-rays are absorbed and their Stokes parameters in detector coordinates, the Pulse Height Amplitude (PHA) of the event (i.e., the X-ray uncalibrated energy), and other parameters related to the photoelectron tracks recorded by the GPDs (see, e.g., \cite{DiMarco22, DiMarco23a}) that are useful for identifying the quality of the tracks in the weighted analysis and the particle background rejection. In addition to scientific data from astrophysical observations, Level-1 data also include possible in-flight calibration runs performed with onboard calibration sources hosted in a Filter and Calibration Wheel \cite{Ferrazzoli20}.

\ixpe Level-2 files include events that occur during periods when the instrument is properly configured for observation, accurately directed at the target, and unobstructed by the Earth or its atmosphere. The level-1 data are processed by the \ixpe instrument pipeline, generating the level-2 data containing: a refined detector position (after removing effects such as dithering, etc.); the energy of the event in terms of Pulse Invariant (PI) channels (ranging from 0 to 374, corresponding to mid-bin energies from 0.02 keV to 14.98 keV with a uniform bin size of 40 eV); the Stokes parameters $Q$ and $U$, corresponding to the initial direction of electron ejection, adjusted for spurious modulation \cite{Rankin22}; the column W\_MOM corresponding to the optimal weight $\alpha_k^{0.75}$ for the weighted analysis previously described. Both detector coordinates and Stokes parameters in Level-2 files are converted into the J2000 tangent plane centered on the target. In this system, the X-axis is aligned with the celestial equator, while the Y-axis is oriented perpendicularly to it and directed towards the north celestial pole. Furthermore, Level-2 data include the \texttt{TIME} column in MET as in Level-1. Since December 18, 2024, the algorithm reported in Ref.~\cite{DiMarco23a} to identify events due to particle background has been applied to Level-2 event files; then the FITS header keyword \texttt{XPFLGBGD=T} is added, and the 8th bit of the STATUS column is set to the value  1 for events identified as background. In the following, a guide for removing these events is provided (see Section~\ref{sec:dc}).

\section{\ixpe response functions}\label{sec:irfs}

\begin{figure}[!ht]
	\centering
	\includegraphics[width=0.4\textwidth]{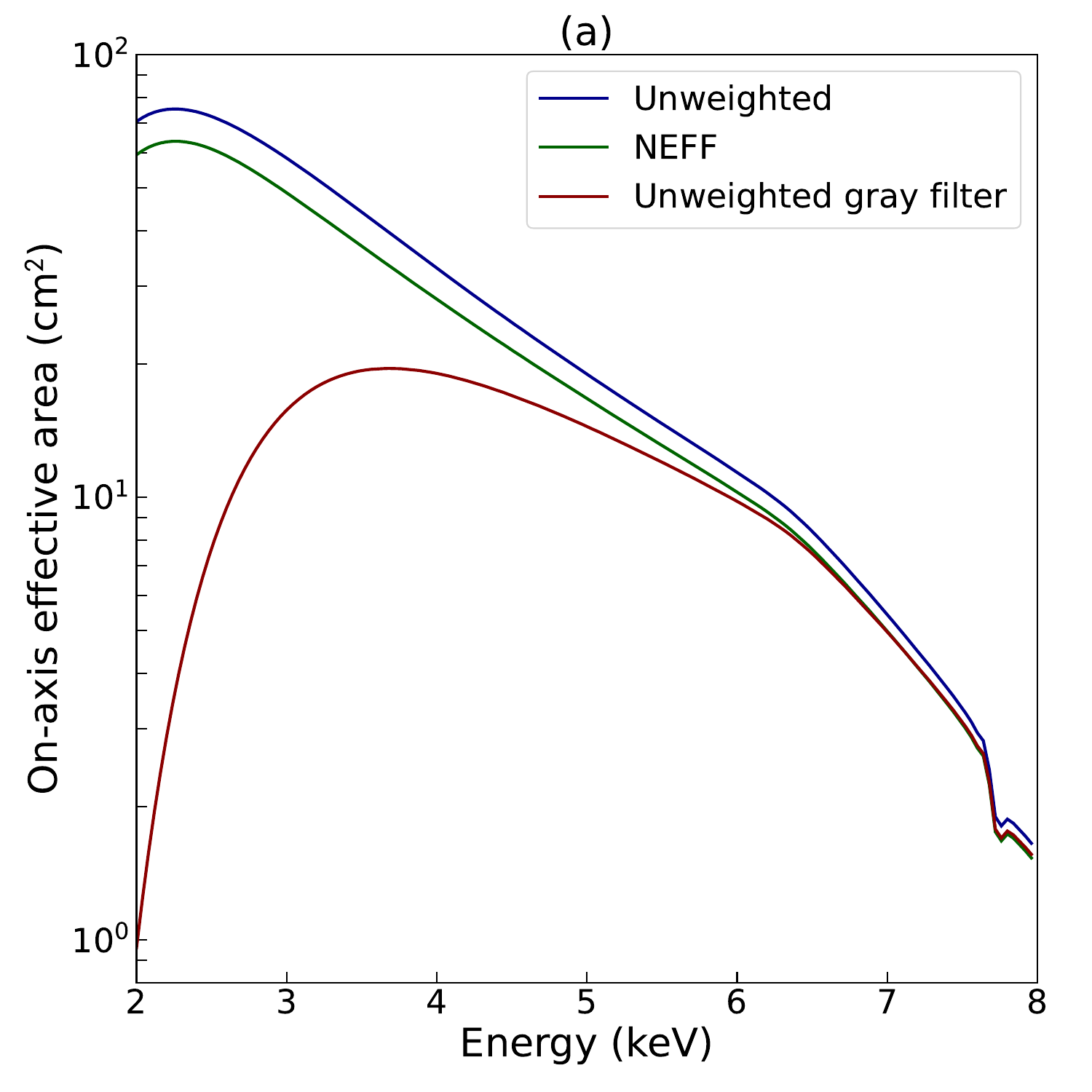}
	\includegraphics[width=0.4\textwidth]{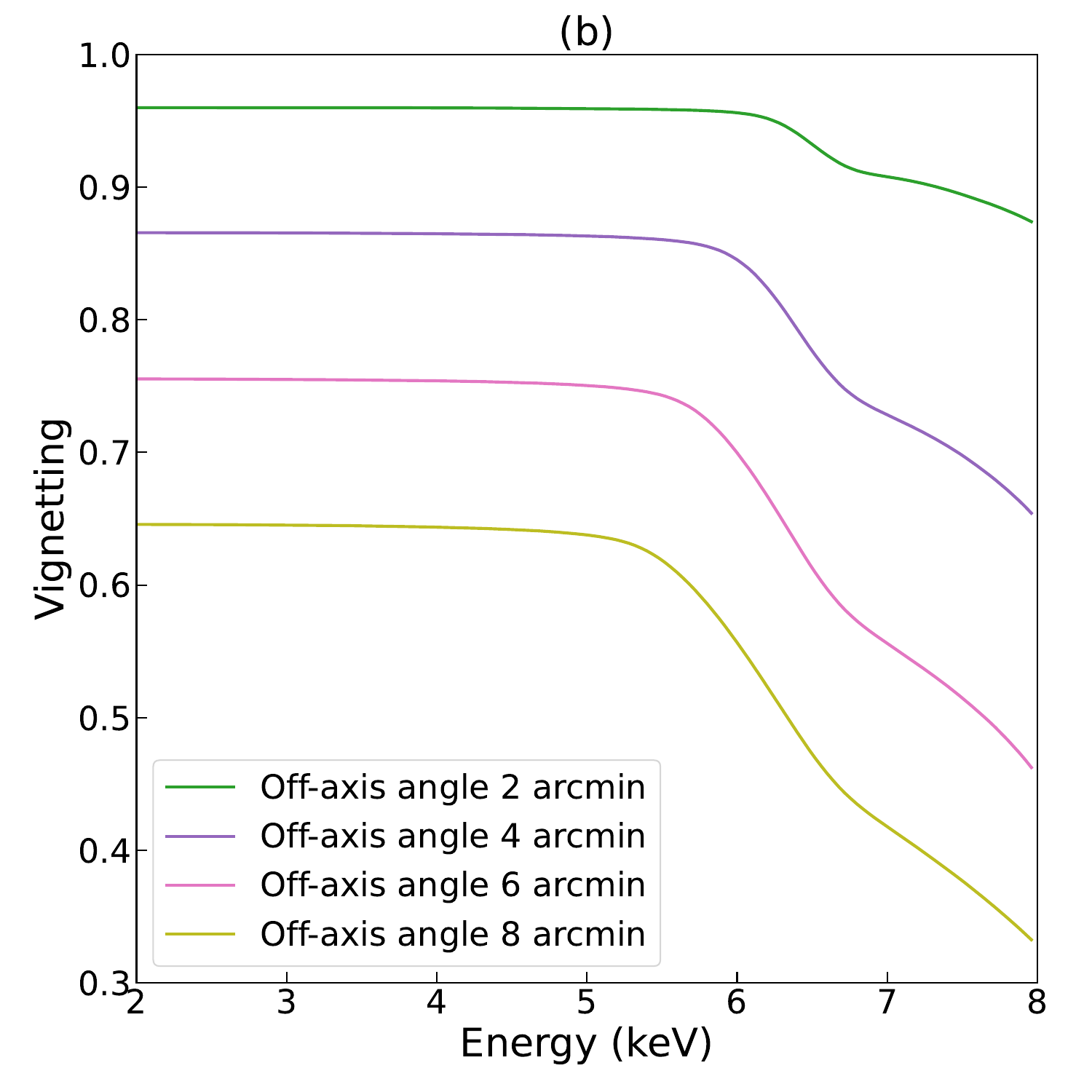}\\
   	\includegraphics[width=0.4\textwidth]{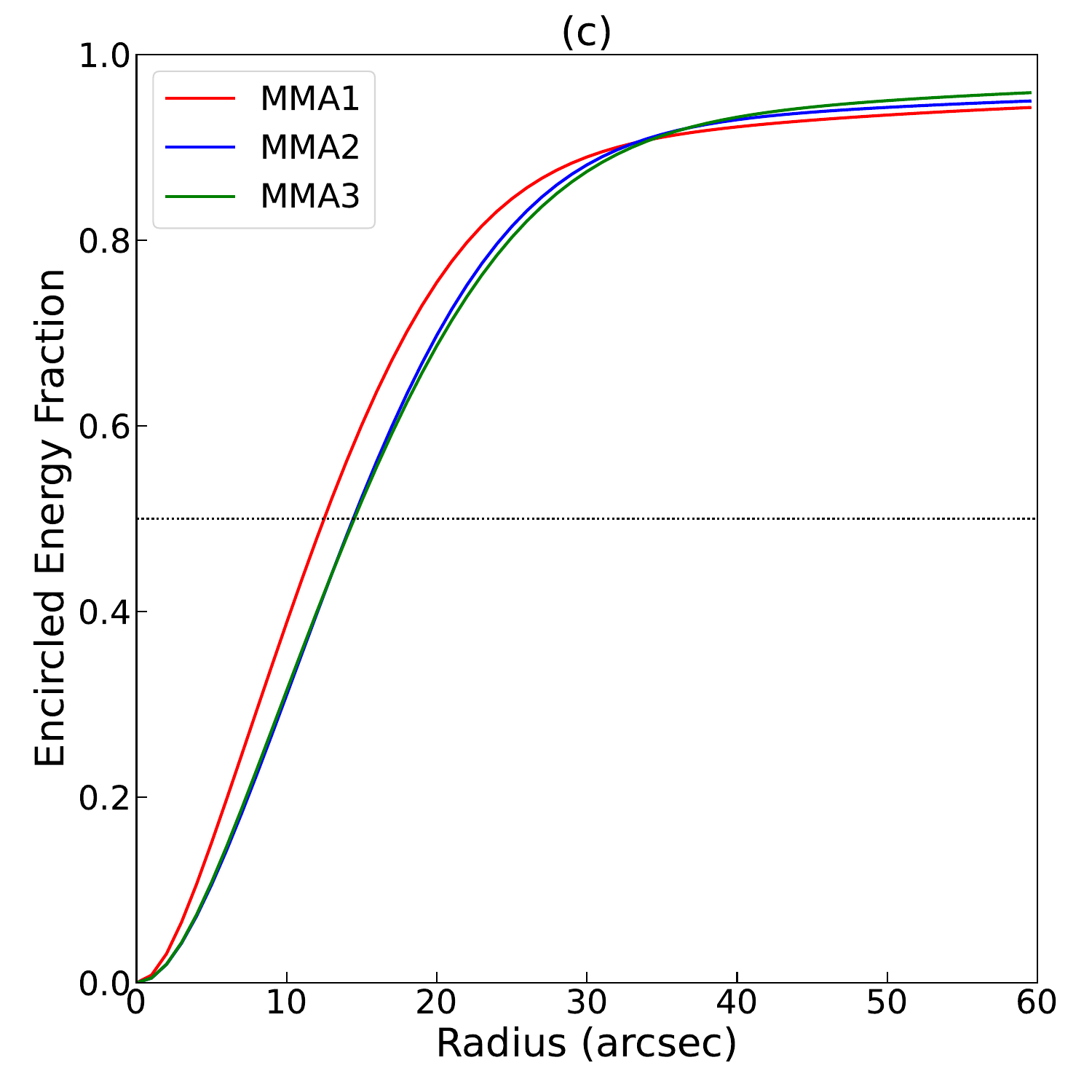}
    \includegraphics[width=0.4\textwidth]{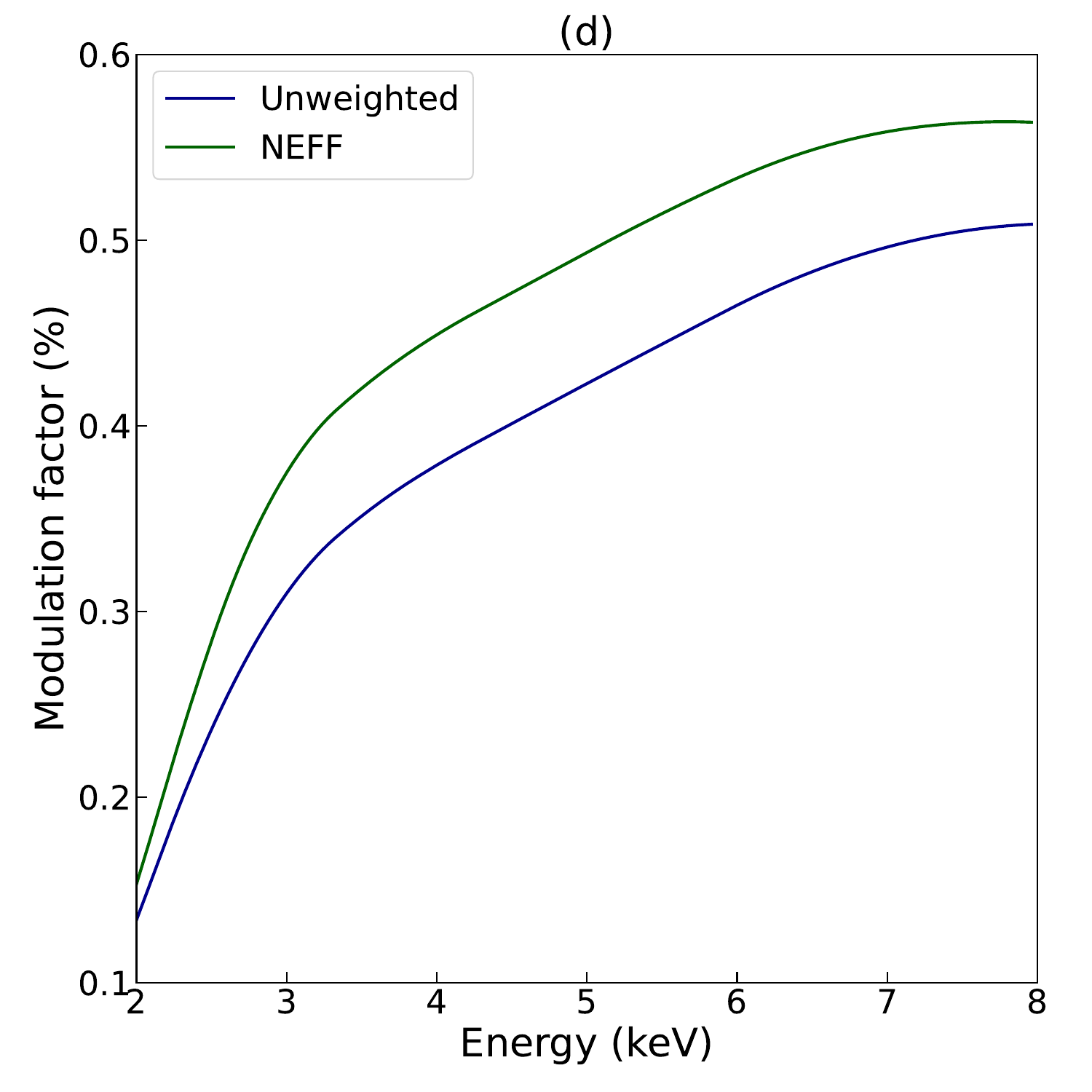}
	\caption{(a) On-axis \ixpe effective area as a function of the energy. The solid lines represent the total effective area for the three telescopes in the unweighted, weighted, and gray-filter versions. (b) Vignetting of the \ixpe optics as a function of energy and off-axis angle. (c) Encircled energy fraction of the three \ixpe telescopes; the dotted line reports the 0.5 value. (d) Modulation factor as a function of the energy for the \ixpe DU1 in the unweighted and weighted versions.}
	\label{fig:irfs}
\end{figure}

The same \ixpe Instrument Response Functions (IRFs) are distributed on HEASARC and included in the software package \textsc{ixpeobssim} with different names and conventions; the latter is an \ixpe contributed software developed by the \ixpe Science Team \cite{Baldini22}. At the time of writing this article, the latest version of IRFs on \textsc{ixpeobssim} is v013, whereas in HEASoft it is the one released on June 29$^{th}$, 2026 (version 20260610). Because of GPD's secular evolution \cite{Baldini21}, the \ixpe IRFs are distributed every 6 months; \textsc{ixpeobssim} provides response files with such a validity time, with the beginning of each interval encoded in the file name (e.g., \texttt{20250701} for the ones valid in the period 1st July - 31st December 2025). While HEASoft provides the ftool \texttt{ixpecalcarf} to automatically select the appropriate set of response files (see later), in \textsc{ixpeobssim} the user must select the correct ones for the observation under analysis. As an example, in Figure~\ref{fig:irfs}a, the total \ixpe on-axis effective area (sum of the three DUs) for the period July 1$^{st}$--December 31$^{st}$, 2025, is reported in the three versions: the unweighted, weighted (\texttt{NEFF}), and gray filter (see later). 

\begin{figure}[!ht]
	\centering
	\includegraphics[width=0.75\textwidth]{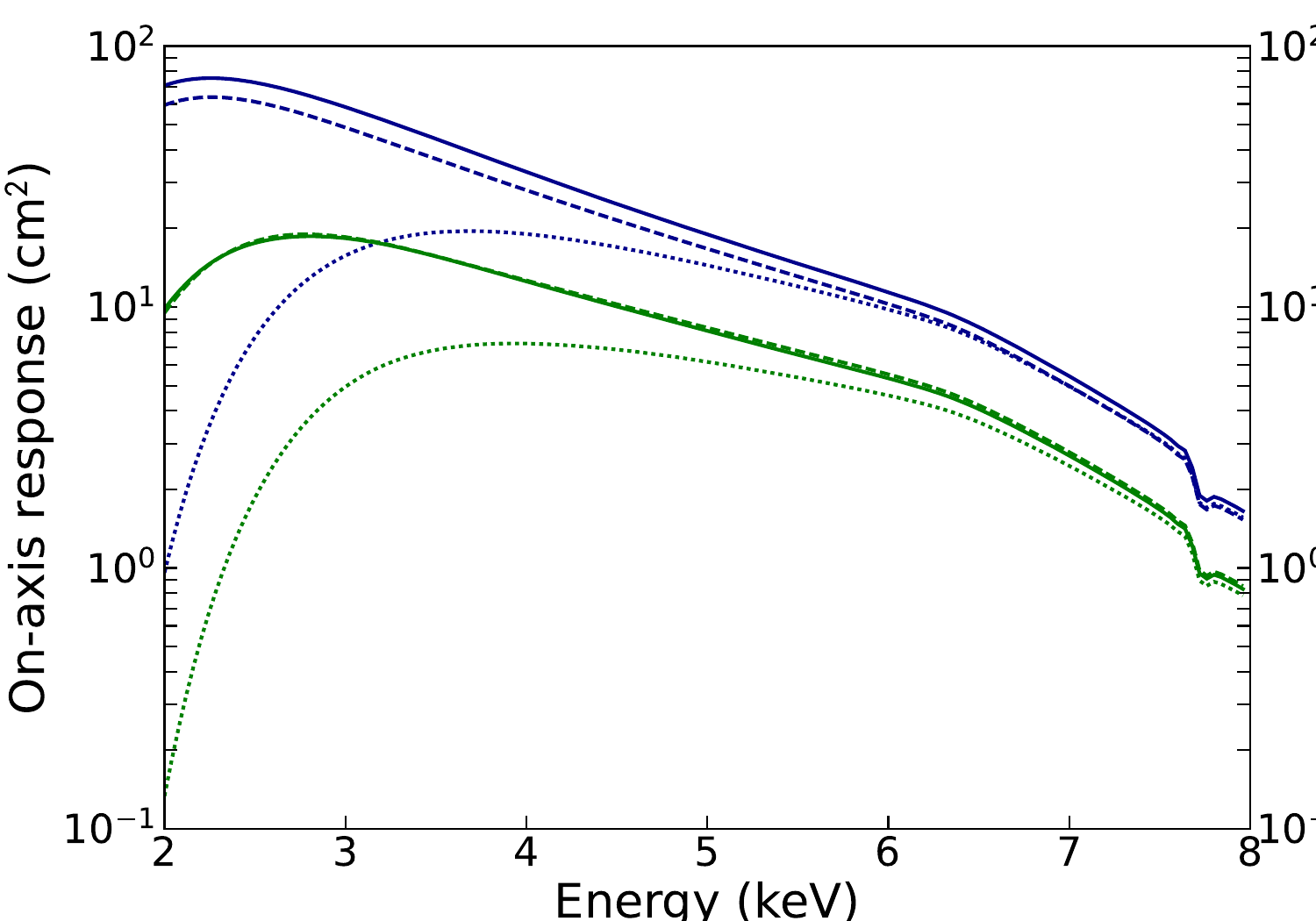}
	\caption{Overall \ixpe response: blue lines report the \texttt{arf}, while the green lines are the \texttt{mrf}; IRFs are reported in the unweighted (solid lines), weighted \texttt{NEFF} (dotted lines), and gray filter (dashed lines) versions.}
	\label{fig:arfs}
\end{figure}

\ixpe analysis requests for six categories of response functions, whose names in some cases are familiar to people that are used to \textsc{xspec} workflow: (i) the effective area (\texttt{.arf}), that is, the equivalent physical area that fully collects radiation, accounting for factors like reflectivity, vignetting, and detector efficiency (see Figure~\ref{fig:irfs}a); (ii) the vignetting, which is a matrix accounting for the loss of image brightness at the edges compared to the center at different energies (see Figure~\ref{fig:irfs}b); (iii) the energy dispersion (\texttt{.rmf}), which is a matrix accounting for the energy resolution of the focal plane detector; (iv) the point-spread function (PSF) describing how \ixpe telescopes blur a single, infinitely small point of light, accounting for optical limitations (Figure~\ref{fig:irfs}c reports the ``Encircled Energy Fraction'', corresponding to the percentage of total light energy from a point source that falls within a specific circular radius); (v) the modulation factor ($\mu$), that is, the response of the \ixpe GPDs to 100\% polarized X-rays \cite{DiMarco22b} (see Figure~\ref{fig:irfs}d); (vi) the modulation response function (\texttt{.mrf}) given by the convolution of the \texttt{arf} and the modulation factor (see Figure~\ref{fig:arfs}).
The \ixpe effective area is obtained by the convolution of the following functions: (i) the MMA effective area \cite{Ramsey2022}; (ii) the transparency of the MMA thermal shield; (iii) the transparency of the detector-unit UV filter \cite{LaMonaca2021}; (iv) the transparency of the GPD window \cite{Baldini21}; (v) the efficiency of the GPD \cite{Baldini21}; the efficiency of the event weighting (in the weighted flavor) \cite{DiMarco22}. The \ixpe energy dispersion, \texttt{rmf}, has been obtained from Monte Carlo simulations. The modulation factor has been calibrated on the ground, as reported in Ref.~\cite{DiMarco22b}, and later monitored in flight \cite{DiMarco2022}. The vignetting function (see Figure~\ref{fig:irfs}b), along with the relative orientation of the three \ixpe DUs, defines the relative exposure across the field of view of the instrument. The \ixpe PSF models are derived for each DU, combining information from one of the early point-source observations and the differences measured during the mirror calibrations \cite{Ramsey2022}. The resulting half-energy width (HEW) obtained as the encircled energy fraction containing 50\% of the events (see Figure~\ref{fig:irfs}c) is ${\sim}$25 arcsec for MMA1 and $\sim$29 arcsec for MMA2 and MMA3. A conservative HEW of $\sim30$ arcsec is considered in spatially resolved analyses of \ixpe data. 
Corrections to the IRFs to account for the vignetting due to dithering, or in the case of extended sources, can be obtained with the \texttt{ixpecalcarf} tool by specifying attitude files included in the \texttt{hk} data folder that contain precise, time-stamped records of a satellite's orientation and pointing direction. As an example, for using \texttt{ixpecalcarf} on a point-like source, the following command line should be used:
\begin{svgraybox}
\texttt{ixpecalcarf evtfile=./event\_l2/ixpe02001099\_det1\_evt2.fits} \\
\texttt{attfiles="./hk/ixpe02001001\_det1\_att.fits,}\\
\texttt{./hk/ixpe02001002\_det1\_att.fits" radius=1.0}\\
\texttt{ specfile="ixpe02001099\_DU1I.pi"}
\end{svgraybox}
Here, observation ID 02001099 consists of two pointings, and two attitude files are required. All the IRFs are defined for each \ixpe DU on a broad 1--12\,keV energy range with an energy binning of 40\,eV. All the IRFs are distributed in three distinct flavors: unweighted, weighted \textsc{NEFF}, and weighted \textsc{SIMPLE} (see Section \ref{sec:specpol} for details). A further set of IRFs for bright sources is the gray-filter set; the latter is an opaque filter used to reduce the low-energy flux from very bright sources, such as \mbox{Sco X-1} \cite{LaMonaca24}. The \ixpe overall spectropolarimetric responses are obtained by combining the elements described above (see Figure~\ref{fig:arfs}), and they are provided by the \texttt{arf} for the $I$ spectrum and by the \texttt{mrf} for the $Q$ and $U$ spectra.


\section{Preliminary data processing}
\label{sec:dc}

As previously reported, \ixpe data are provided by HEASARC, and they can contain a README file in the observation ID folder indicating possible issues or warnings\footnote{We remark that at the time of writing, several observations performed during the \ixpe GO2 include a README file reporting that at TIME=2.614352714230000E+08, some pixels failed on DU2; DU2 data distribution restarted after this issue was fixed, but comparison of the polarimetric results obtained from each DU is encouraged to verify that DU2 provides consistent results.}. As an example, observations performed before May 2022 reported possible problems due to off-axis and energy scale that the user had to correct by hand using proper tools. 
The preliminary data processing presented here provides suggestions that may be needed only for some observations, but they must be verified before data extraction and analysis to avoid issues in the polarimetric measurement. The following subsections will describe the different procedures in the same order as they need to be performed, following the workflow for the processing.

\subsection*{Image alignment}

\begin{figure}[!bh]
\centering
\includegraphics[width=0.4\textwidth]{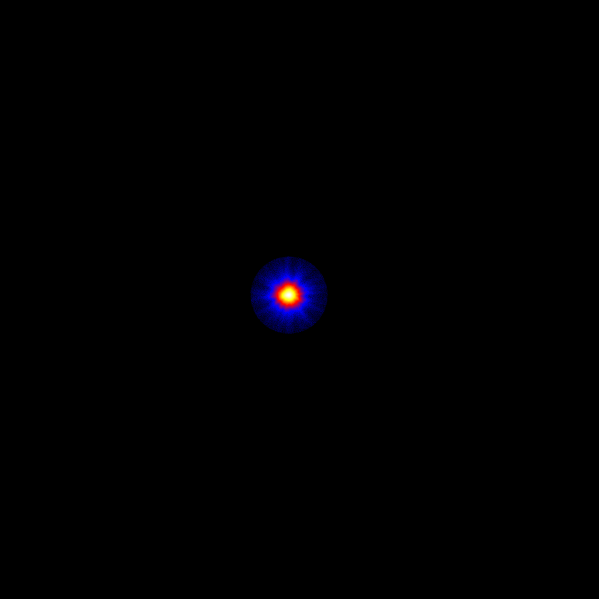}
\includegraphics[width=0.4\textwidth]{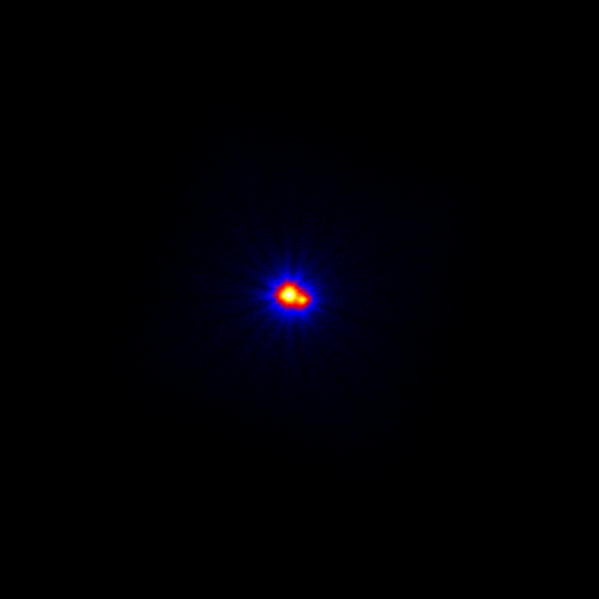}
\caption{Example of a good (left) and a bad (right) alignment between two observational segments of the source \mbox{4U~1820$-$303}.}
\label{fig:align}   
\end{figure}

To avoid effects arising from mixing different emission regions in extended sources or from possible systematics, such as ``polarization leakage'' \cite{Bucciantini2023}, the images collected from the three \ixpe DUs and/or different observation IDs have to be well aligned. In Figure~\ref{fig:align}-right, an example of a bad merging of two observation segments for the source \mbox{4U~1820$-$303} is shown. 
To align them, the merged Level-2 files should be divided into the different pointings. Then, the image position in the detector's coordinate frame for each pointing has to be determined. At this point, it is possible to modify these coordinates in pixel units, as reported in the parameters \texttt{TCRPX7} and \texttt{TCRPX8} of each file, using the tool \texttt{fmodhead} available in HEASoft.

\subsection*{Source and background selections}

In the presence of dim sources, with a source rate, $S$, comparable to that of the background, $B$, the contribution from the latter must be accounted for. In this case, the measured polarization degree, $P_{meas}$, is affected by dilution with respect to the intrinsic polarization of the source, $P_{src}$, because the background is expected to be not polarized \cite{DiMarco23a}:
\begin{equation}
    P_{meas} = P_{src} \left( 1 + \frac{B}{S} \right)^{-1}.
\end{equation}
The corresponding MDP to account for this, when a measurement lasts for a certain time, $T$, is obtained by modifying Equation~\ref{eq:mdp} into \cite{DiMarco23a, Weisskopf2010}:
\begin{equation}
\textrm{MDP}_{99}\approx \frac{4.29}{\mu S} \sqrt{\frac{S+B}{T}}.
\end{equation}
It is evident that the background contribution reduces the significance. Thus, reducing the background is essential to achieving the maximum sensitivity and avoiding dilution effects. In the case of bright point-like sources, the background is usually assumed to be negligible, and the source can be spatially selected thanks to \ixpe imaging capabilities. For fainter sources (although the background region can also be spatially selected) and for extended sources, even the brightest ones, the background must be considered and handled appropriately. 

\begin{figure}[!ht]
	\centering
	\includegraphics[width=0.65\textwidth]{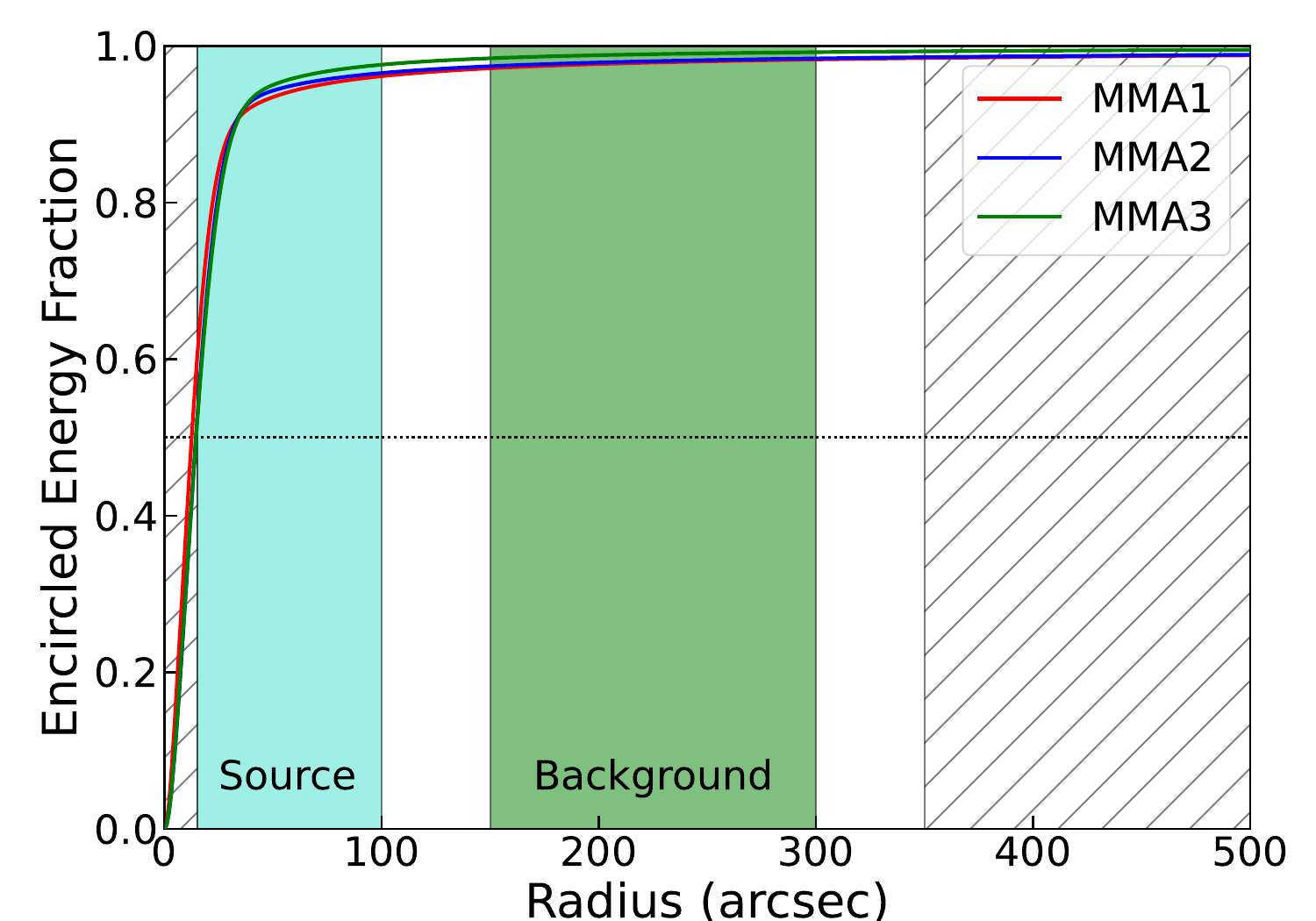}
	\caption{\ixpe imaging selection radii for source (light blue) and background (green). Hashed regions are due to values smaller than HEW or large enough to introduce systematic effects on polarization due to detector-edge effects.}
	\label{fig:selections}
\end{figure}

\ixpe observes a single source at a time using the ``point-and-stare'' observing strategy, and the source is optimally centered within the three detectors' sensitive areas. Each DU has a field of view of $\sim$13 arcmin \cite{Weisskopf2022}, but the DUs' clocking and the dithering pattern reduce the common field of view to $\sim$9 arcmin. A guide that accounts for the polarimetric response depending on the spatial selection of source and background is provided in Ref.~\cite{DiMarco23a} and is summarized here. The point-like sources can be spatially selected in a circular region at least equal to the \ixpe PSF HEW, which is ${\sim}$30 arcsec (see Figure~\ref{fig:irfs}c), similarly extended sources can be studied by applying spatial selections larger than $\sim$30 arcsec. This is a conservative approach to obtaining polarization in spatially independent bins, but techniques for studying polarization in smaller regions have been developed (see, e.g., Ref.~\cite{Bucciantini2023b}). For the background selection, the study of the polarization as a function of the radius showed that the polarization of the source is detected up to ${\sim}$100 arcsec from the center of the image, while at radii larger than ${\sim}$350 arcsec a significant polarization begins to be detected due to geometrical effects of the detector; in fact, because of dithering, the source at these radii begins to touch the edges of the detector \cite{DiMarco23a}. Thus, the best choice for the background selection is an annular region with an inner radius of 150 arcsec and an outer radius of up to 300 arcsec. In Figure~\ref{fig:selections}, a summary of the spatial selections provided in Ref.~\cite{DiMarco23a} is given.

In conclusion, \ixpe detectors use imaging to select background events, resulting in the optimal selection of an annular region with an inner radius of 150 arcsec and an outer radius of 300 arcsec. For the source selection, a circular region with a radius of ${>}$15 arcsec can be used, with a maximum radius of up to 100 arcsec. For the source selection, when the source is faint, smaller radii are preferable to reduce background contamination.

\subsection*{Removal of solar flares}

\begin{figure}[!ht]
	\centering
	\includegraphics[width=0.85\textwidth]{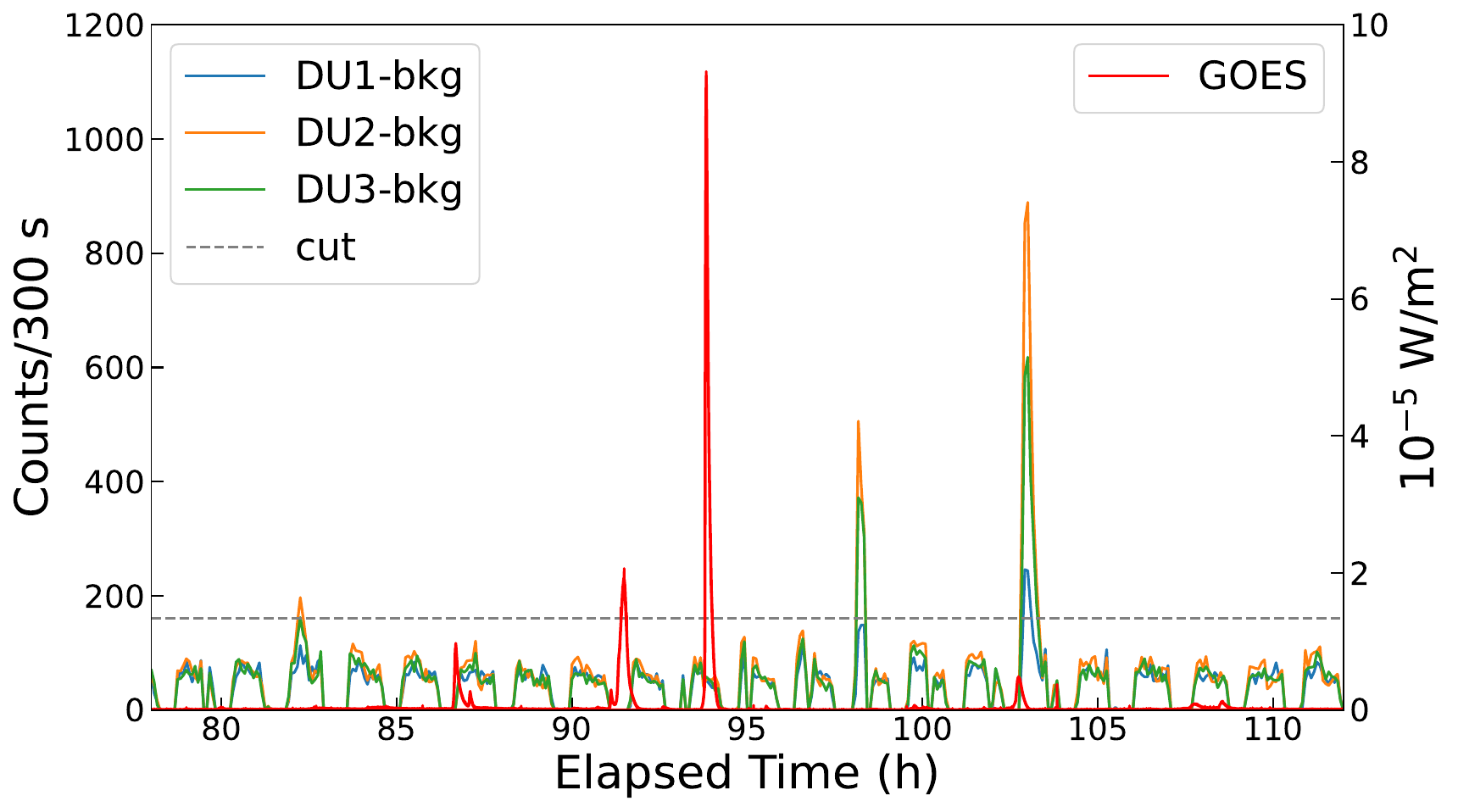}
	\caption{\ixpe light curve in time bins of 300\,s when the background region is selected in DU1 (blue), DU2 (orange), and DU3 (green). The \ixpe light curve shows periodic (${\sim}$96.6\,min) gaps in the data due to Earth occultation and passage in the South Atlantic Anomaly. The GOES data in W/m$^2$ (right axis) are reported in red. Three spikes are visible in both the GOES and the \ixpe light curves; these differ in the time scale due to the different orbits of the observatories, and \ixpe data show differences in the three DUs due to different exposure to the Sun; a flux cut to remove these time intervals is also reported in a dashed gray line. See text.}
	\label{fig:flares}
\end{figure}

Some \ixpe observations, such as Vela~PWN \cite{Xie22}, \mbox{Vela~X$-$1} \cite{Forsblom2023}, \mbox{3C~58} \cite{Bucciantini25}, \mbox{NGC~2110} \cite{Chakraborty2025} and \mbox{RCW~86} \cite{Silvestri26}, were affected by intense solar flare activity. When solar flares are detected by \ixpe, they impact the results if not properly handled \cite{Bucciantini25,Chakraborty2025,Silvestri26}, producing a fake polarized signal and contributing to the spectra. Some X-ray sources also show intrinsic peculiar variability, such as flux variations related to orbital and spin periods or due to type I X-ray bursts; thanks to a more stable count rate, it is easier to identify the time intervals where the solar-related events are present in the background region. As an example, in Figure~\ref{fig:flares}, the \ixpe light curves from the three DUs in the background-selection region are superimposed on the one from the Geostationary Operational Environmental Satellites (GOES) during the \ixpe observation of \mbox{Vela~X$-$1}.
It is possible to observe in the \ixpe background region that the count rate exhibits flares related to those detected by GOES. As a further interesting feature, this kind of event is not uniform in the three \ixpe DUs because they are not equally exposed to the Sun. In particular, DU2, being more exposed to the Sun, can be affected by stronger contamination from solar activity. The standard background rejection (see later) can mitigate the effect of this increased background, but it is not sufficient to remove all these events; background subtraction can help in this, but for bright sources (as explained in the following), it should be avoided, and for extended sources, when the background selection is not trivial, this is difficult to achieve. A good practice is to remove time intervals affected by solar-related events. To achieve this, it is important to perform the time filtering before background rejection, as the latter can partially remove solar-related events, making time selection more difficult.

Thus, to better identify the solar-related events, the following procedure is suggested:
\begin{itemize}
    \item extract for each \ixpe DU the events in the background region using the prescription previously reported;
    \item compare the light curves of the three \ixpe DUs:  A pronounced flux increase in DU2 relative to DU1 and DU3 indicates the presence of solar-related events (see Figure~\ref{fig:flares});
    \item using the DU2 average count rate ($R_2$) and its standard deviation ($\sigma_{R_2}$), a flux selection on the time (see Figure~\ref{fig:flares}) is performed to identify affected time intervals: $R(t) {>} R_2+ 3\times\sigma_{R2}$.
\end{itemize}
In the end, the time selection provided by DU2 can be used on the three \ixpe DUs to remove solar-related events.

A characterization of these solar-related events is provided in Appendix~A of Ref.~\cite{Chakraborty2025}. In particular, these events are due to illumination from the side of the \ixpe spacecraft; beyond producing different fluxes across the three DUs, their directionality yields a polarization of ${\sim}20$\% \cite{Chakraborty2025, Silvestri26}. The latter is not the polarization of solar flares but the product of a geometrical effect. Moreover, the effects of solar-related events are also prominent in the \ixpe spectra, resulting in the presence of Si and Al fluorescence lines \cite{Chakraborty2025} when spacecraft daytime spectra are compared with nighttime spectra that are not affected by these events \cite{Chakraborty2025,Silvestri26}. Since version 32,  \textsc{ixpeobssim} includes a specific tool for subtracting the contribution to polarization from solar flares (S. Silvestri et al., in prep.); this approach does not provide a ``cleaned'' file, but accounts for the presence of solar events in the products of \textsc{ixpeobssim}, as presented in Section~\ref{sec:polobs}. 

\subsection*{Background rejection and subtraction}

Even in the absence of polarization, particles contributing to the instrumental background can induce a signal that mimics X-ray polarization. Distinguishing between the X-ray polarimetric signal and such a fake one is very difficult for the \ixpe detectors because particle identification is not straightforward. Consequently, hereafter, this fake signal is considered a background-induced polarization.

In Ref.~\cite{DiMarco23a}, a detailed study comparing the X-ray signal with other detected signals was performed to characterize and reject potential background contributions. Such a comparison permitted the identification of morphological properties of the imaged photoelectric tracks in the \ixpe GPDs, allowing the separation of X-ray-like and particle-like signals. In particular, background events due to particles tend to produce longer tracks and lower charge collection per pixel \cite{DiMarco23a, Xie21} in IXPE GPDs because of their virtually constant (and minimal) energy loss; moreover, in these tracks, a higher frequency of active pixels in the track border is found. In light of this, Ref.~\cite{DiMarco23a} studied all the available \ixpe track properties to find the optimal ones for background-source decoupling; the useful parameters for this aim are the number of pixels in the track, the energy/charge fraction in the primary cluster over the total charge recorded for the single event, and the number of boundary pixels. All these parameters can be found in the \ixpe Level-1 data but not in the Level-2 ones; thus, to associate events included in Level-1 with the ones included in Level-2, only the \texttt{TIME} parameter is available. Because of this, background rejection must be performed before any time correction. 
In particular, in Ref.~\cite{DiMarco23a} the following strategy has been developed:
\begin{itemize}
    \item Number of pixels: X-rays produce tracks that are longer at higher energy (more active pixels), whereas background events, although producing longer tracks than X-rays, do not show such a strong correlation in the \ixpe energy band. Because of this, an energy-dependent threshold to identify particle-background events can be applied: 
    \begin{equation}
        \texttt{NUM\_PIX}>70+30\times E\textrm{ (keV)}.
    \end{equation}
    \item Energy fraction: GPD readout tracks can feature more clusters of pixels, but the event reconstruction algorithm considers only the biggest one, defined as the primary/main cluster. Because the energy release from background particles is almost constant along longer tracks and is typically split into more clusters, the main cluster from particles stores a smaller fraction of the energy/charge compared with X-rays. The identification of particles on the basis of this parameter is provided by:
    \begin{eqnarray}
        \texttt{EVT\_FRA}& > &1 \\
        \texttt{EVT\_FRA}& < & 0.8 \times \left( 1-e^{-\frac{E\textrm{ (keV)}+0.25}{1.1}}\right)+0.004 \times E\textrm{ (keV)}. \nonumber
    \end{eqnarray}
    \item Border pixels: particle-induced background events can produce longer tracks and more clusters, resulting in tracks with a higher probability of having active pixels at the edges; thus, to identify background:
    \begin{equation}
        \texttt{TRK\_BORD} > 1.
    \end{equation}
\end{itemize}
By combining these three conditions, background identification is achieved.

The background rejection tool is available at the following link: \href{https://github.com/aledimarco/IXPE-background}{https://github.com\\/aledimarco/IXPE-background}. For the \ixpe data processed after December 18$^{th}$, 2024, the background tool is applied to Level-2 data that report a \texttt{FLGBGD} and tag the particle background, setting the 8$^{th}$ bin in the \texttt{STATUS} to 1 for that event. It means that at the present time, for data processed after December 18$^{th}$, 2024, the background rejection can be applied with the following command: 
\begin{svgraybox}
\texttt{ftselect ./event\_l2/ixpe04000099\_det1\_evt2\_v??.fits} \\ \texttt{./event\_l2/ixpe04000099\_det1\_evt2\_rej.fits} \\
\texttt{status.NE.bxxxxxxx1xxxxxxxx}
\end{svgraybox}
\noindent
while for data prior to December 18$^{th}$, 2024, the \texttt{filter\_background.py} has to be used with the following command:
\begin{svgraybox}
\texttt{filter\_background.py ixpe04000099\_det1\_evt2.fits} \\ 
\texttt{ixpe04000001\_det1\_evt1.fits ixpe04000002\_det1\_evt1.fits}
\end{svgraybox}
\noindent
where \texttt{ixpe04000099\_det1\_evt2.fits} is the Level-2 originating from the \texttt{ixpe04000001\_det1\_evt1.fits} and \texttt{ixpe04000002\_det1\_evt1.fits} Level-1 files. A \texttt{--output} option can allow for the default \texttt{rej} value, producing a Level-2 file including only source events; \texttt{bkg} value, producing a Level-2 file including only events identified as background; and \texttt{tag} option, producing a new column in the Level-2 file containing, for each event, 1 for events identified as source and 0 for those identified as background.

This rejection approach has a negligible impact on the response matrices and can remove up to $\sim$40\% of background events; the \ixpe background is measured at 0.003 cps/arcmin$^2$ \cite{DiMarco23a}. This rejection method is more effective for faint sources, as brighter sources have the background-selection region dominated by source events due to the \ixpe PSF \cite{DiMarco23a}. Consequently, three potential background treatments may be executed:
\begin{itemize}
    \item[i] Bright sources (rate${>}$2 cps/arcmin$^2$), where background is negligible, and rejection is possible but ineffective, and subtraction from the same field of view should be avoided;
    \item[ii] Intermediate sources, where the background rejection is effective, but the background region remains dominated by source events; thus, subtraction in the analysis should be avoided.
    \item[iii] Faint sources (rate${<}$1 cps/arcmin$^2$) where background rejection is effective and the remaining background contribution should be subtracted in the analysis.
\end{itemize}
\vspace{0.5cm}
After checking that the images are well aligned, that possible solar events are properly removed, and that background rejection is applied, one can select source and background regions in the final data for the next steps. The following provides guidance for data extraction and analysis; the subsequent sections assume that all previous steps have been successfully completed. During \ixpe GO2, the DU2 experienced an anomaly; as a result, a new approach for background rejection was developed. In the Appendix, the new rejection parameters are reported after the anomaly, and the rejection is suggested to be applied to DU2 data also for bright sources because of the increased number of noisy pixels.

\section{Model-independent polarimetric analysis}\label{sec:polobs}

The previous sections described how to prepare and select data for the analysis. At this point, it is possible to obtain a measurement of the source polarization without any spectral modeling or assumption by using the \textsc{ixpeobssim} software \cite{Baldini22} or the new \ixpe contributed software \texttt{ixpe\_protractor}.

\begin{figure}[!b]
	\centering
	\includegraphics[height=0.35\textwidth]{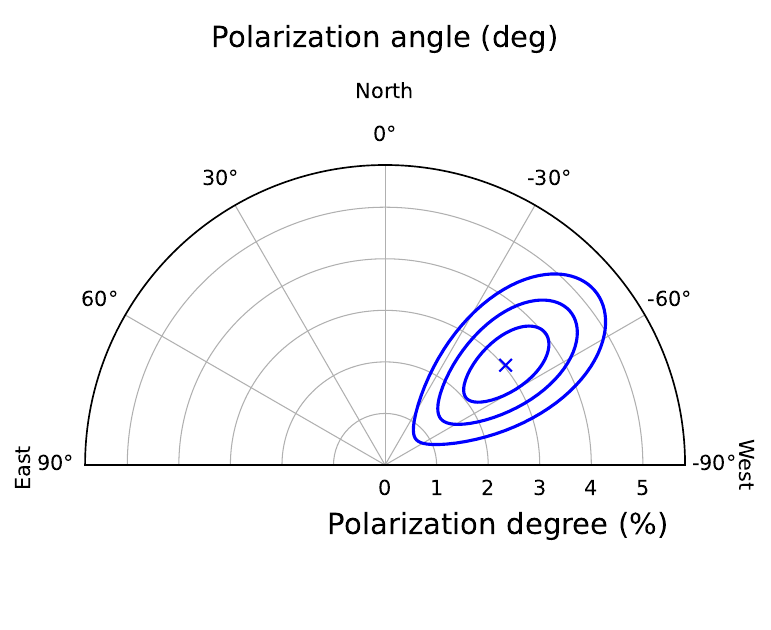}
	\includegraphics[height=0.35\textwidth]{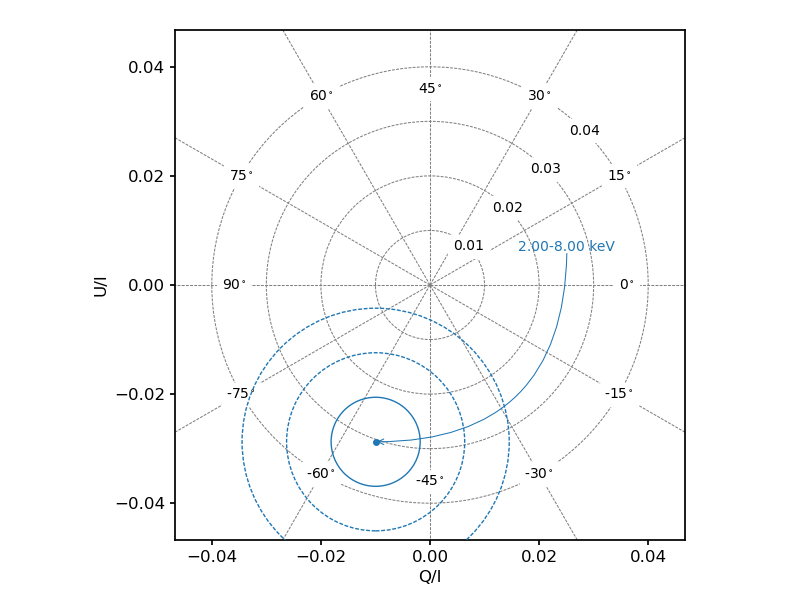}
	\caption{Left: example of a protractor plot provided by the \texttt{ixpe\_protractor} script; lines from inner to outer report regions at 68\%, 95\% and 97\% CL. Right: example of a polarization measurement obtained with the \texttt{PCUBE} algorithm in the \texttt{xpbin} tool of \textsc{ixpeobssim}; the lines report the uncertainties at 1$\sigma$, 2$\sigma$ and 3$\sigma$ in the normalized $Q$ and $U$ plane. Both examples are obtained for the \ixpe observation ID 01002501.}
	\label{fig:mod_ind}
\end{figure}

\texttt{ixpe\_protractor} is a Python script using the \texttt{ixpepolarization} tool in HEASoft and the \ixpe CALDB; it provides the measurement of polarization in a certain energy bin and the corresponding protractor plot in polar coordinates (see, for example, Figure~\ref{fig:mod_ind}-left).
The script requires a region file for the spatial selection of the source (\texttt{src.reg}) that can be obtained by SAOImageDS9\footnote{\href{https://sites.google.com/cfa.harvard.edu/saoimageds9}{https://sites.google.com/cfa.harvard.edu/saoimageds9}} as described in Section~\ref{sec:dc}, and an energy interval (in this example, the nominal 2--8\,keV \ixpe energy band):
\begin{svgraybox}
\texttt{python3 ixpe\_protractor.py ixpe01002501\_det?\_evt2\_rej.fits --region=src.reg --eband='2.0 8.0'}    
\end{svgraybox}
\noindent
The script will also provide the PD, PA, $I$, $Q$, $U$, and MDP values for each \ixpe DU and for the combined data. For bright point-like sources, this will provide a preliminary measurement of the polarization. If background subtraction is needed, the \texttt{ixpe\_protractor} script should be used on data from the background region to obtain the background Stokes parameters. Thus, the following relation is used to estimate the intrinsic source polarization:
\begin{eqnarray}
        I_{src} & = & I_{m} - \zeta I_{bkg} , \nonumber \\
        Q_{src} & = & Q_{m} - \zeta Q_{bkg} , \\
        U_{src} & = & U_{m} - \zeta U_{bkg} , \nonumber
    \end{eqnarray}
where the subscripts $m$ and $bkg$ indicate the measured polarization in the source and background regions, respectively, while $\zeta$ is a scaling factor:
\begin{equation}
    \zeta = \frac{A_{src}T_{src}}{A_{bkg}T_{bkg}},
\end{equation}
where $A_{src}$ and $A_{bkg}$ are the areas for the source and background spatial selections, $T_{src}$ and $T_{bkg}$ are the exposure times (in general, the same exposure times will be obtained if selections are performed on the same observation).

The same analysis can be performed with the \textsc{ixpeobssim} software by using the \texttt{xpbin} tool. The set of analysis tools distributed in \textsc{ixpeobssim} includes the following:
\begin{itemize}
    \item \texttt{xpbin} produces binned event lists using different algorithms and provides polarimetric results;
    \item \texttt{xpphase} that calculates the phase of a periodic source based on its ephemeris;
    \item  \texttt{xpselect} filtering event lists based on energy, time, phase, etc.;
    \item \texttt{xpbinview} that allows for visualizing the results from \texttt{xpbin}.
\end{itemize}
\texttt{xpbin} provides different algorithms to bin the data products (the complete list and its explanation are provided on the dedicated website\footnote{\href{https://ixpeobssim.readthedocs.io/en/latest/}{https://ixpeobssim.readthedocs.io/en/latest/}} and in Ref.~\cite{Baldini22}); here, only a simple analysis is presented to obtain polarization for a point-like source in the source region in a certain energy interval using the ``polarization cube'' (\texttt{PCUBE}) algorithm.
To generate the products from the \texttt{PCUBE} algorithm in \texttt{ixpeobssim}, one simply has to execute the \texttt{xpbin} script with the option designated for the algorithm argument:
\begin{svgraybox}
\texttt{xpbin ixpe01002501\_det*\_evt2\_rej\_src.fits --algorithm PCUBE --ebins 1 --irfname ixpe:obssim20220702:v013}    
\end{svgraybox}
\noindent
where the name of the response matrices is also explicit and reports the start date of the validity period, and the \texttt{ebin} option indicates a single energy bin in the 2--8\,keV energy interval. The output will be a set of files, one for each DU, with the polarimetric results. These files can be read by using the \texttt{xpbinview} tool:
\begin{svgraybox}
\texttt{xpbinview ixpe01002501\_det*\_evt2\_rej\_src\_pcube.fits}
\end{svgraybox}
\noindent
\texttt{xpbinview} reads the \texttt{xpbin} products, providing as the output the polarization parameters and a plot, as in Figure~\ref{fig:mod_ind}-right. For background subtraction, the same considerations for \texttt{ixpe\_protractor} apply.

Since April 2026, \textsc{ixpeobssim} includes the new tool \texttt{xpsun.py}, which can subtract the contribution from solar flares in the polarimetric analysis (S. Silvestri et al., in prep.). This is obtained by running:
\begin{svgraybox}
    \texttt{xpsun --l2files ixpe01002501\_det*\_evt2\_rej.fits}
\end{svgraybox}
\noindent
This script provides two files containing the data split into two parts: \texttt{insun}, including data acquired when the detector is illuminated by the Sun; \texttt{ineclipse}, including data during the period when the Sun is not visible to the detector. These two datasets can be used in \texttt{xpbin} to obtain polarimetric results after subtracting the contribution due to solar flares:
\begin{svgraybox}
\texttt{xpbin ixpe01002501\_det*\_evt2\_rej\_src.fits --algorithm
PCUBE --ebins 1 --irfname ixpe:obssim20220702:v013 --insun ixpe01002501\_det*\_evt2\_rej\_insun.fits --ineclipse ixpe01002501\_det*\_evt2\_rej\_ineclipse.fits}
\end{svgraybox}
\noindent
The \texttt{PCUBE} file in the output will provide the net source polarization after accounting for solar flares. This method is not providing a new file after removal of time intervals affected by solar events but is providing the output of \texttt{xpbin} with its final products; even if this method has the limitation of not providing a ``cleaned’’ file, in some cases it is the only viable one to account for the presence of solar flares, especially for extended sources where background selections are challenging.

\section{Spectro-Polarimetric analysis}\label{sec:specpol}

\textsc{xspec} \cite{Arnaud96} enables simultaneous fitting of the Stokes parameter spectra in a spectro-polarimetric analysis \cite{Strohmayer2017}, including linear polarization. The latter follows from the fact that Stokes parameters are additive and can be treated as flux quantities: $I$ represents the source flux, whereas $Q$ and $U$ encode the source's polarization. Given that $I$, $Q$, and $U$ fully characterize the source emission, they can be used to assign different polarizations to different spectral components in this model-dependent analysis.

\textsc{xspec} for the spectro-polarimetric analysis needs $I$, $Q$ and $U$ spectra; for this purpose, the \texttt{XFLT0001} keyword of the SPECTRUM extension has been introduced with the value \texttt{0} associated with $I$ spectra, \texttt{1} associated with $Q$ spectra, and \texttt{2} associated with $U$ spectra. To properly extract the data useful for the spectro-polarimetric analysis, one can use the standard ftool \texttt{extractor} to obtain the $I$, $Q$ and $U$ spectra with 40\,eV energy bins in the different available flavors. An alternative approach is based on the use of \texttt{xpbin} in \textsc{ixpeobssim} using the algorithms \texttt{PHA1}, \texttt{PHA1Q} and \texttt{PHA1U} for the $I$, $Q$ and $U$ spectra, respectively. 

As anticipated in Section~\ref{sec:irfs}, \ixpe analysis can use three different approaches: the \textsc{UNWEIGHTED}, the weighted \textsc{SIMPLE} and the weighted \textsc{NEFF}. The \textsc{UNWEIGHTED} case, when data are extracted for spectro-polarimetric analysis, provides:
\begin{eqnarray*}
    I = \frac{N}{T} \hspace{1cm} & \sigma_I = \frac{\sqrt{N}}{T}, \nonumber \\
    Q = \frac{1}{T} \sum_k q_{k} \hspace{1cm} & \sigma_Q \approx \frac{1}{T}\sqrt{2N}, \\
    U = \frac{1}{T} \sum_k u_{k} \hspace{1cm} & \sigma_U \approx \frac{1}{T}\sqrt{2N}.
\end{eqnarray*}
When the \textsc{SIMPLE} approach is used for data analysis, given $N_{eff}=I^2/W_2$, the extraction will provide in each energy bin:
\begin{eqnarray*}
    I = \frac{1}{T} \sum_k w_k \hspace{1cm} & \sigma_I = \frac{1}{T} \sqrt{I}, \nonumber \\
    Q = \frac{1}{T} \sum_k w_k q_{k} \hspace{1cm} & \sigma_Q \approx \frac{1}{T}\sqrt{2W_2}, \\
    U = \frac{1}{T} \sum_k w_k u_{k} \hspace{1cm} & \sigma_U \approx \frac{1}{T}\sqrt{2W_2},
\end{eqnarray*}
assuming $w_k=1$, the unweighted case is obtained from the same equations. The \textsc{NEFF} provides:
\begin{eqnarray*}
    I = \frac{N_{eff}}{T} \hspace{1cm} & \sigma_I = \frac{\sqrt{N_{eff}}}{T}, \nonumber \\
    Q = \frac{N_{eff}}{TI} \sum_k w_k q_{k} \hspace{1cm} & \sigma_Q = \frac{N_{eff}}{IT} \sqrt{\sum_k (w_k q_{k})^2}, \\
    U = \frac{N_{eff}}{TI} \sum_k w_k u_{k} \hspace{1cm} & \sigma_U = \frac{N_{eff}}{IT} \sqrt{\sum_k (w_k u_{k})^2}.
\end{eqnarray*}

A recently distributed script named \texttt{ixpestartx}\footnote{\href{https://heasarc.gsfc.nasa.gov/docs/ixpe/analysis/contributed/ixpestartx.html}{https://heasarc.gsfc.nasa.gov/docs/ixpe/analysis/contributed/ixpestartx.html}} can be used to simplify the extraction of the \ixpe Stokes spectra for analysis using \textsc{xspec}. Given the observation ID, the chosen weighting scheme, the source and (optional) background extraction regions in SAOImageDS9 files format, the script will pick the proper directories for Level~2 event files and related attitude files to extract $I$, $Q$, and $U$ spectra and to use \texttt{ixpecalcarf} to build the specific \texttt{arf} and \texttt{mrf} for each spectrum and an \textsc{xspec} \texttt{xcm} file to be directly used in \textsc{xspec} for uploading the data and pointing to the proper IRFs. The \texttt{ixpestartx} can be used as in the following example:
\begin{svgraybox}
    \texttt{./ixpestartx 01002501 neff /path/to/src.reg /path/to/bkg.reg}
\end{svgraybox}

\begin{figure}[!bth]
	\centering
	\includegraphics[width=0.49\textwidth]{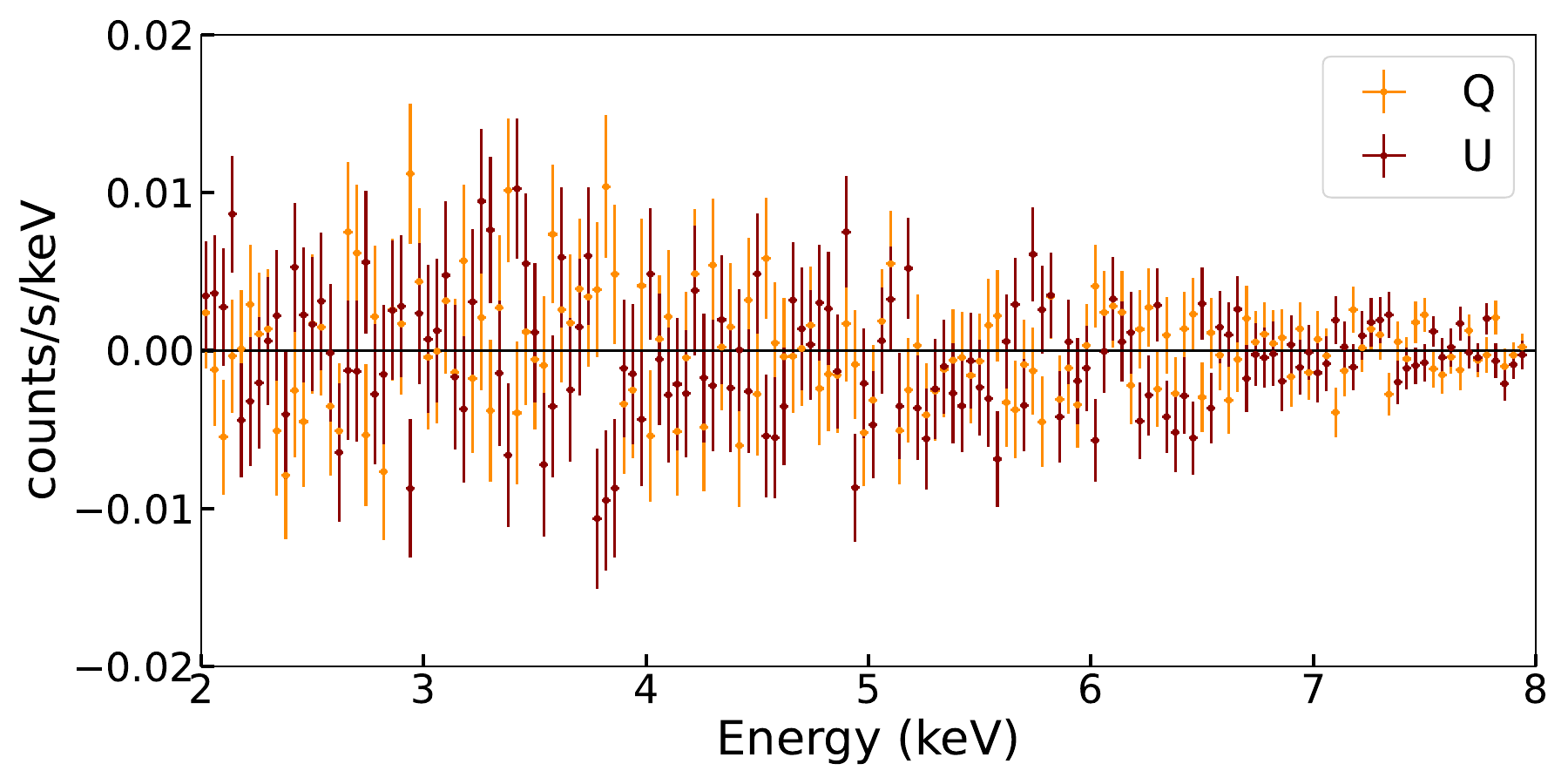}
	\includegraphics[width=0.49\textwidth]{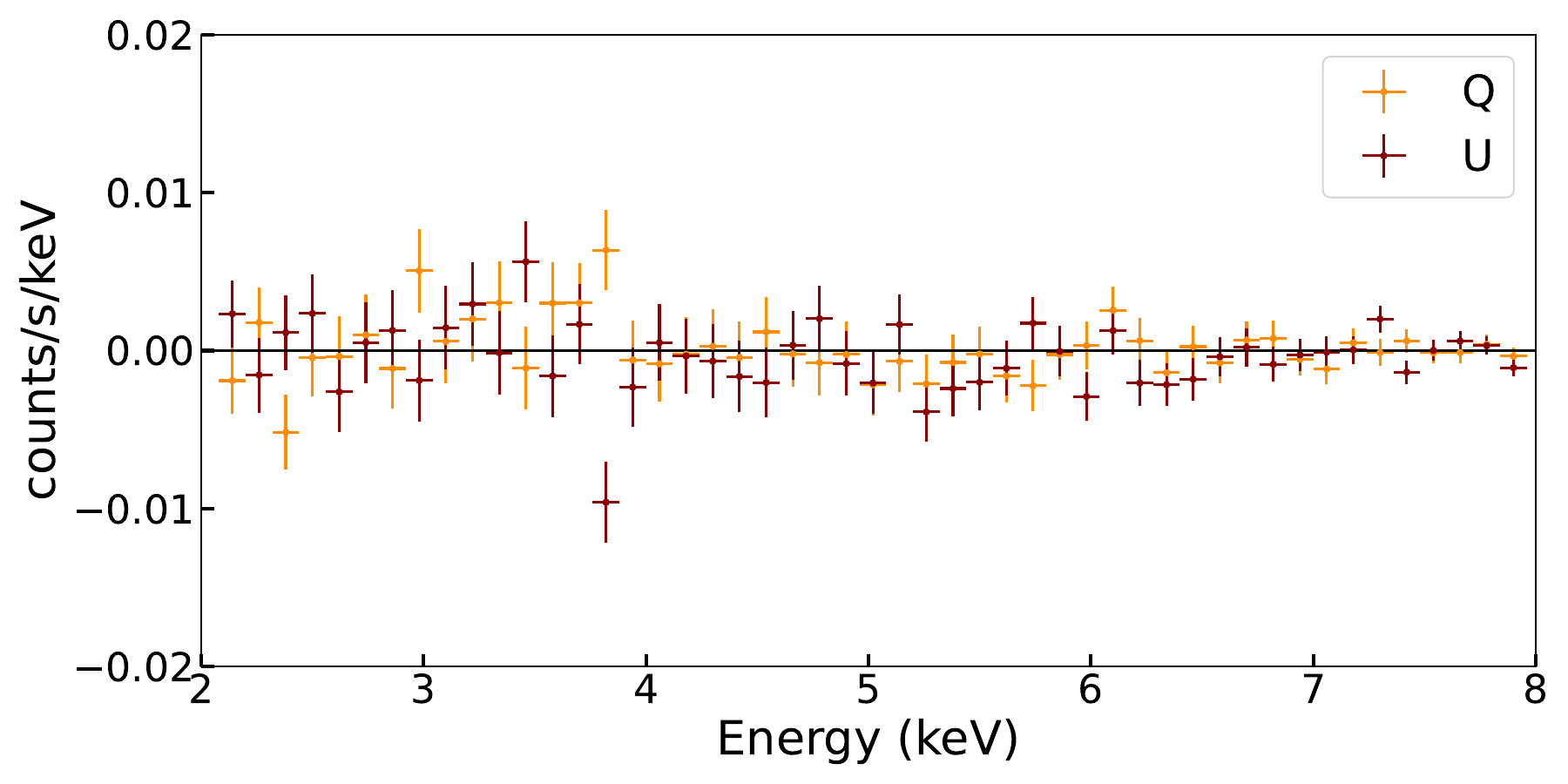}
	\caption{Comparison for the $Q$ (orange) and $U$ (red) spectra in the case of no rebinning (left) and after a constant rebinning (right) for the \ixpe observation ID 01002501.}
	\label{fig:rebin}
\end{figure}

After extraction, the $I$, $Q$, and $U$ spectra need to be rebinned. The $I$ spectrum is a ``standard'' spectrum, and the rebinning follows the typical suggestions for $\chi^2$ statistics with a minimum number of 30 counts, for example, using the following command:
\begin{svgraybox}
    \texttt{grppha ixpe\_det3\_src\_NEF\_I.pha ixpe\_det3\_src\_NEF\_I\_rb.pha comm="group min 30 \& chkey ANCRFILE ixpe\_det3\_NEF.arf \& chkey RESPFILE ixpe\_d3\_20170101\_alpha075\_02.rmf \& exit"}
\end{svgraybox}
For the $Q$ and $U$ spectra, which can also have negative values, a binning based on the minimum number of counts is not helpful; furthermore, as shown in Figure~\ref{fig:rebin}-left, the uncertainties are typically larger than the mean values of a bin. 
To handle this, a constant binning of 3 or 5 bins (120 or 200\,eV, respectively) to reduce the uncertainties should be used:
\begin{svgraybox}
    \texttt{grppha ixpe\_det3\_src\_NEF\_U.pha ixpe\_det3\_src\_NEF\_U\_rb.pha comm="group 1 375 3 \& chkey ANCRFILE ixpe\_det3\_NEF.mrf \& chkey RESPFILE ixpe\_d3\_20170101\_alpha075\_02.rmf \& exit"}
\end{svgraybox}
After extracting and rebinning the spectra, it is possible to load them into \textsc{xspec} with the following command:
\begin{svgraybox}
        \texttt{data 1:1 ixpe\_det1\_src\_NEF\_I\_rb.pha \\
    data 1:2 ixpe\_det1\_src\_NEF\_Q\_rb.pha \\
    data 1:3 ixpe\_det1\_src\_NEF\_U\_rb.pha}
\end{svgraybox}
\noindent
where only one DU is reported, but similarly the remaining two must be loaded. $I$, $Q$ and $U$ spectra for each DU are associated with the same source.
This spectro-polarimetric analysis is performed similarly to standard spectral analyses in \textsc{xspec}. The user defines a model describing the source spectrum, and the latter is convoluted with the IRFs to obtain the measured expected spectrum; this is given in the case of the Stokes parameters by:
\begin{eqnarray}
    I_m(E) &=& \int_{E'} I_s(E')\epsilon(E')R(E',E)dE', \nonumber \\
    U_m(E) &=& \int_{E'} U_s(E') \mu(E')\epsilon(E')R(E',E) dE' ,\nonumber \\
    Q_m(E) &=& \int_{E'} Q_s(E') \mu(E')\epsilon(E')R(E',E) dE',\nonumber
\end{eqnarray}
where the $m$ subscript identifies the expected measured fluxes, $s$ the theoretical source spectrum, $R(E'E)$ the \texttt{rmf} that redistributes the events from an initial energy $E'$ to an energy $E$ accounting for the energy resolution of the detector, $\epsilon(E')$ the \texttt{arf} and $\mu(E')\epsilon(E')$ the \texttt{mrf}; all of them are described in Section~\ref{sec:irfs}. Then, these expected spectra are compared with the real data $I$, $Q$ and $U$ from the instrument to optimize a statistical index; typically, the adopted method is the $\chi^2$ minimization:
\begin{equation}
    \chi^2 = \frac{\sum_i (I_i-I_{m,i})^2}{\sum_i\sigma_{I,i}^2},
\end{equation}
where $i$ is the index running over all the considered energy bins and $\sigma_{I, i}$ is the uncertainty associated with that bin. The same equation is applied to the $Q$ and $U$ spectra.

Currently, three polarimetric models are officially available in \textsc{xspec} that provide phenomenological approaches to describing the energy dependence of X-ray polarization in spectropolarimetric analyses: \texttt{polconst}, \texttt{pollin}, and \texttt{polpow}. They are multiplicative components that modify the Stokes parameters of the underlying emission model, allowing a simultaneous fit of the intensity spectrum and the polarization observables (i.e., the Stokes $Q$ and $U$ spectra). 

The \texttt{polconst} model assumes that both the polarization degree and polarization angle are constant over the entire energy range, providing the simplest description of the data. 

The \texttt{pollin} model introduces a linear energy dependence on both the polarization degree and the polarization angle. In this case, the polarization degree and angle are described as follows:
\begin{eqnarray}
  \textrm{PD}(E) & = & A_1+A_{\rm slope}(E[\textrm{keV}]-1\,\textrm{keV}),\\
   \textrm{PA}(E) & = & \psi_1+\psi_{\rm slope}(E[\textrm{keV}]-1\,\textrm{keV})\,, \nonumber
\end{eqnarray}
where $A_{\rm slope}$ and $\psi_{\rm slope}$ are the slopes of the linear trends, and $A_1$ and $\psi_1$ are the values of the degree and angle of polarization at the energy of 1 keV. 

The \texttt{polpow} model assumes a power-law dependence on energy for polarization degree and angle:
\begin{eqnarray}
    \textrm{PD}(E) & = & A_{\rm norm}E[\textrm{keV}]^{-A_{\rm index}},\\
    \textrm{PA}(E) & = & \psi_{\rm norm}E[\textrm{keV}]^{-\psi_{\rm index}}, \nonumber
\end{eqnarray}
where $A_{\rm norm}$, $\psi_{\rm norm}$, $A_{\rm index}$ and $\psi_{\rm index}$ are the norm and the index of the power-law dependence on the energy of the angle and degree of polarization, respectively.

A possible spectropolarimetric model could be the following: 
\begin{center}
    \texttt{constant*TBabs*powerlaw*polconst},\\
\end{center}
where a cross-normalization constant is introduced to account for possible differences in the \texttt{arf} of the three DUs. For bright sources, it is possible that the three DUs show a bad fit $\chi^2$ for the $I$ spectral model; in that case, the use of \texttt{gain fit} can help to fix this issue related to calibration uncertainties.

\begin{figure}[!htb]
	\centering
	\includegraphics[width=0.45\textwidth]{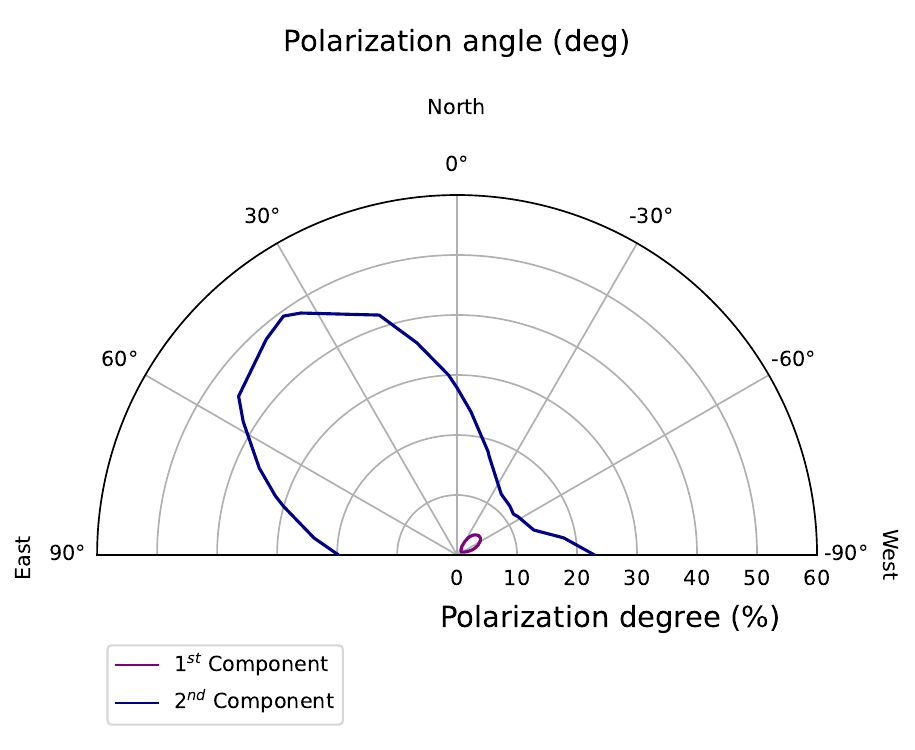}
	\includegraphics[width=0.45\textwidth]{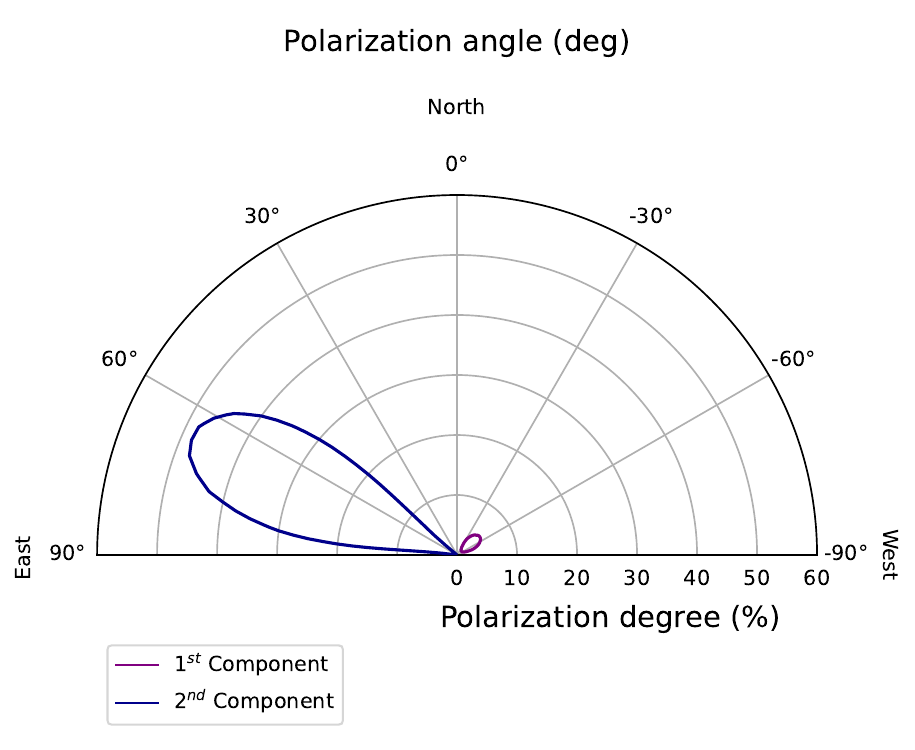}
	\caption{Results for two-component spectropolarimetric analysis as obtained by \texttt{steppar} when properly performed (left) or not properly performed (right).}
	\label{fig:steppar}
\end{figure}

The polarimetric results obtained by \textsc{xspec}, as discussed in Section~\ref{sec:pol}, should be presented in a protractor plot that accounts for the correlation between PD and PA. This is achieved with the \texttt{steppar} command, using $\Delta\chi^2$ values for 2 parameters at the specified confidence levels. In some analyses, when multiple spectral components are present, \textsc{xspec} allows for associating a different polarization with each spectral component, such as:
\begin{center}
   \texttt{const*TBabs*(polconst*diskbb+polconst*bbodyrad)}.\\
\end{center}
In this case, \texttt{steppar} should be run on each pair of PD and PA parameters to obtain the protractor plots as shown in Figure~\ref{fig:steppar}-left.
In some cases, authors want to freeze the PA of one component with respect to the other one; in this scenario, \texttt{steppar} will not allow for a run on PD and PA of the component with the PA frozen, and some authors simply use the other PA in \texttt{steppar}, obtaining a wrong estimation of the protractor plots, as in Figure~\ref{fig:steppar}-right. The panels in Figure~\ref{fig:steppar} have the same PA for the first component and compatible PAs for the second one, but in the right panel, the second component has a PA frozen at 90$^\circ$ apart from the PA of the first one. The left panel contour shows that the second component is unconstrained,  whereas the right panel shows for the same component a contour with a strange shape and narrower than expected. This is due to an improperly estimated uncertainty. The reason for this wrong uncertainty resides in the fact that the two components start to be correlated. As an example, if they are parallel, the two PDs will sum, while for orthogonal PAs, they will subtract; both scenarios produce a correlation in the estimate of the two PD values, making them strongly correlated, and at this point, a $\Delta\chi^2$ value for 2 parameters underestimates the uncertainty because the correlation is between 3 parameters: PD$_1$, PA$_1$ and PD$_2$.

\section{Conclusion}

In this chapter, the \ixpe data analysis techniques are described in detail, covering how to properly process the data and both model-dependent and model-independent approaches, using the software available at the time of writing. The chapter summarizes the estimation of Stokes parameters on an event-by-event basis, including the correction for instrumental systematic effects and the use of weights to improve sensitivity. The formats and structure of \ixpe data products are explained, along with instructions for retrieving them and for handling the presence of background and solar-related events. The \ixpe response functions are introduced, along with the available data-extraction approaches: unweighted and weighted (both \texttt{SIMPLE} and \texttt{NEFF}). In the end, spectropolarimetric analysis is introduced with the available polarimetric models in \textsc{xspec}. In the appendix, new background prescriptions after an anomaly occurred at DU2 during the \ixpe GO2 are reported.



\begin{acknowledgement}
I thank the IXPE team for their work and the authors of the various tools and methods presented here, which help users analyze IXPE data every day. I thank the organizers of the Francesco Lucchin INAF PhD School for encouraging me to compile the information now collected in this chapter.  I also thank Fabio La Monaca, Dawoon E. Kim (INAF-IAPS), Anna Bobrikova, Sofia Forsblom (Turku University), Kuan Liu (Guanxi University), and Stefano Silvestri (INFN-Pisa) for their helpful suggestions and comments, which helped me improve the final version of this chapter.
\end{acknowledgement}


\section*{Appendix - Background rejection after DU2 anomaly}\label{sec:du2anomaly}
\addcontentsline{toc}{section}{Appendix}

During the \ixpe GO2 period, at the Mission Elapsed Time of 2.614352714E+08\,s, some pixels failed on DU2, and its data distribution has been temporarily suspended. After this, a campaign of observations to study and characterize such an anomaly was conducted, and one of the results is an updated background-handling procedure, included in the dedicated software\footnote{\href{https://github.com/aledimarco/IXPE-background}{https://github.com/aledimarco/IXPE-background}.} starting from version 3.2. To obtain a new rejection strategy, the same approach as in Ref.~\cite{DiMarco23a} has been followed, and new optimal cuts have been determined for the three useful parameters: \texttt{NUM\_PIX}, \texttt{EVT\_FRA}, and \texttt{TRK\_BORD}. In the following, the observation ID 04252301 is considered as a case study. The background and source populations, selected from the images, were compared for each parameter to tune the new background-rejection strategy. 

\begin{figure}[!htb]
\centering
    \includegraphics[width=0.45\textwidth]{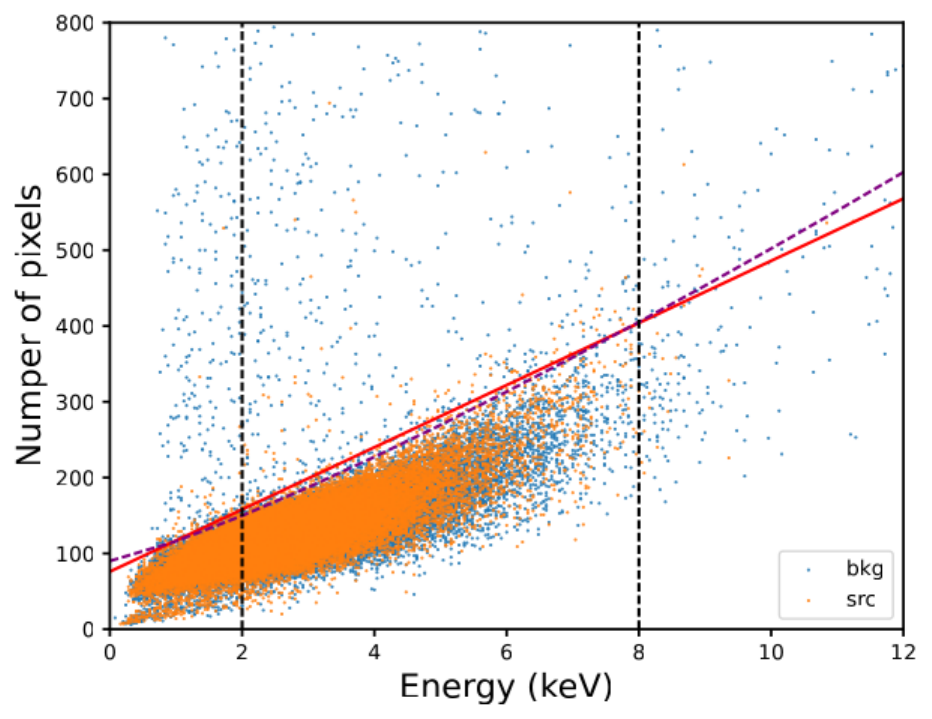}
    \includegraphics[width=0.45\textwidth]{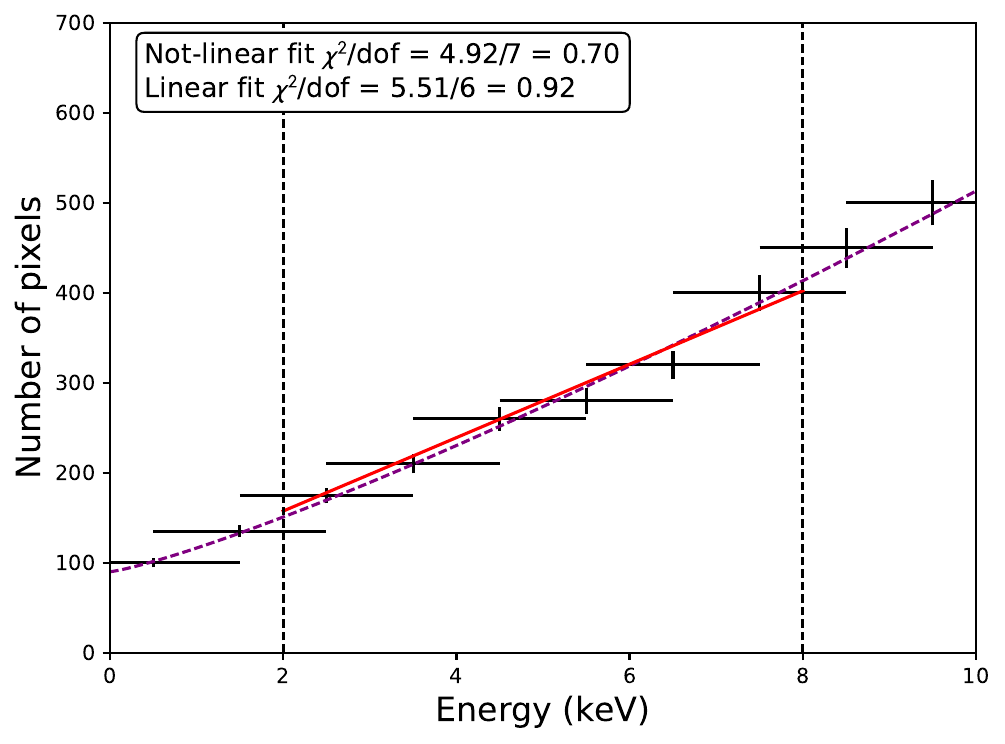}   
	\caption{Left: a scatter plot showing the number of pixels as a function of the energy for the source (orange) and the background (light blue) populations; the vertical black dashed lines show the \ixpe nominal energy band, and the purple dashed line is the threshold on the number of pixels as a function of energy following Equation~\ref{eq:numpix}, while the red one is obtained for a linear cut. Right: Threshold on the number of pixels --- for which the source events above it are negligible --- as a function of energy (black points); the purple dashed line shows a best fit in the 0--10 keV energy band for the Equation~\ref{eq:numpix}, and the red line is a simplified linear function valid in the 2--8 keV \ixpe nominal energy band. }
	\label{fig:num_pix}
\end{figure}

\begin{figure}[!htb]
\centering
	\includegraphics[width=0.45\textwidth]{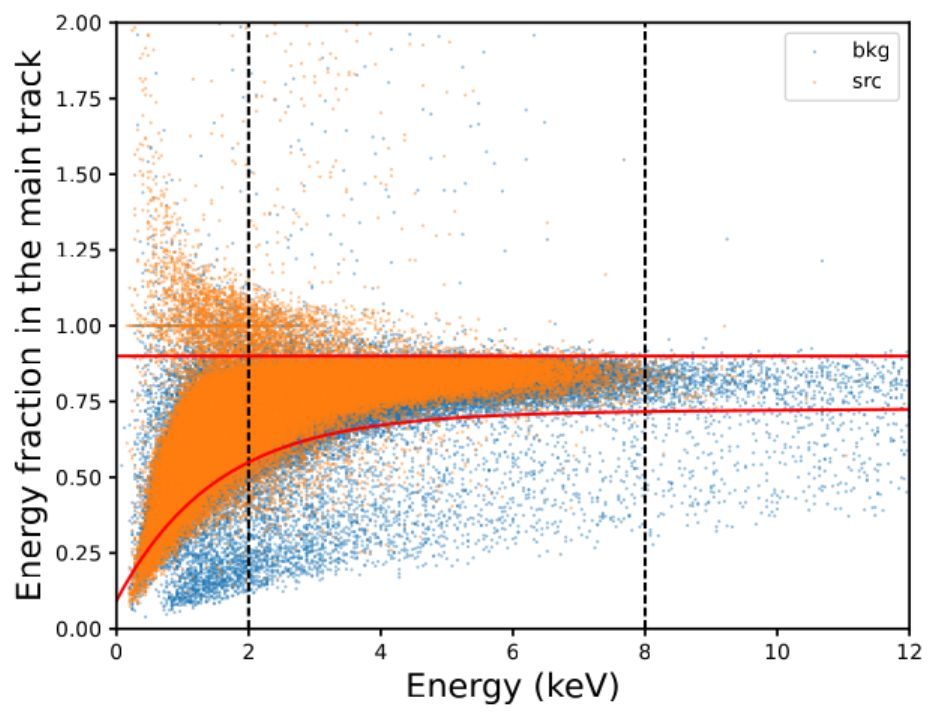}
	\includegraphics[width=0.45\textwidth]{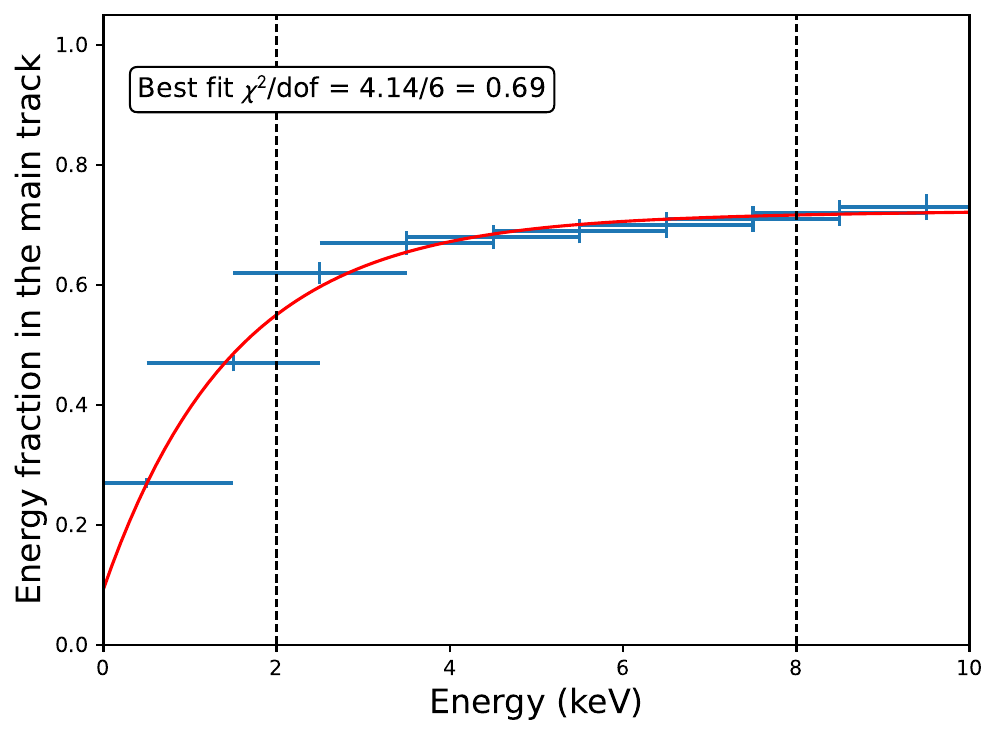}
	\caption{Left: a scatter plot as in Figure~\ref{fig:num_pix}, with red lines indicating the energy-fraction thresholds in the main cluster at 0.9 and from the energy-dependent function given in Equation~\ref{eq:evt_fra}. Right: threshold on the energy fraction as a function of energy, below which source events are negligible (light blue points); the red line shows a best fit in the 0--10 keV energy band.}
	\label{fig:evt_fra}
\end{figure}

The number of pixels after the anomaly shows an average increase at every energy (see Figure~\ref{fig:num_pix}-left), and using the same function adopted by \cite{DiMarco23a}:
\begin{equation}\label{eq:numpix}
\textrm{Number of pixels} < k_a + k_b \times \left( \frac{\textrm{E}}{\textrm{1 keV}} \right)^{k_c},
\end{equation}
the best fit ($\chi^2/dof = 0.92$) were obtained (see Figure~\ref{fig:num_pix}-right):
\begin{itemize}
    \item $k_a = 90\pm6$,
    \item $k_b = 26 \pm 4$,
    \item $k_c = 1.20 \pm 0.08$.
\end{itemize}
In the \ixpe 2--8\,keV nominal energy band, this equation can be simplified with a linear function ($\chi^2/dof = 0.7$), as shown in Figure~\ref{fig:num_pix}-right:
\begin{equation}
\textrm{Number of pixels} = 76(5) + 40.8(1.5) \times E.
\end{equation}
The energy fraction (\texttt{EVT\_FRA}) takes into account the ratio between the energy/charge collected in the main cluster and the one collected in all the detected clusters, and after this anomaly, it results in being slightly reduced on average (see Figure~\ref{fig:evt_fra}-left).
Furthermore, after the DU2 anomaly, the energy fraction as a function of energy shows a population of low-energy events with higher \texttt{EVT\_FRA} values. Although there are not that many events of this kind, these seem to depend on both the spatial region on the detector surface and the counting rate. For this reason, it is better to remove them from any \ixpe observation, regardless of the brightness. On the other side, the threshold for removing background events also changed, and using the same function as in Ref.~\cite{DiMarco23a}, the new parameters can be obtained (see Figure~\ref{fig:evt_fra}-right):
\begin{equation}
\textrm{Fraction of energy} = k_a \times \left[ 1 + e^{-\frac{E+k_b}{k_c}} \right] + k_d \times E
\label{eq:evt_fra}
\end{equation}
with best-fit ($\chi^2/dof = 0.69$) parameters:
\begin{itemize}
	\item $k_a = (0.71 \pm 0.05)$,
	\item $k_b = (0.21 \pm 0.07) \textrm{ keV}$,
	\item $k_c = 1.5 \pm 0.2 \textrm{ keV}$,
	\item $k_d = (0.0012 \pm 0.0006) \textrm{ keV}^{-1}$.
\end{itemize}

Regarding the last parameter, \texttt{TRK\_BORD}, as reported in Ref.~\cite{DiMarco23a}, no energy-dependent behavior is observed; however, after the anomaly, an increase in events with large \texttt{TRK\_BORD} values is observed, likely due to more noisy pixels. Thus, the old condition is too strong to apply, introducing systematic effects in the measured polarization; as a result, the condition for background rejection was relaxed, and the new selection is given by the cut for border pixels ${>}5$.

In conclusion, after the anomaly, new cuts in the background rejection have to be adopted, and they are included in the \ixpe CALDB and in the software \texttt{filter\_background.py} starting by version 3.2. Aiming to remove the population of events with a high energy fraction (${>}$0.9), the rejection procedure should be applied to DU2 data in any case, also for bright sources. 



\bibliographystyle{utphys}
\bibliography{ref.bib}


\end{document}